\documentclass[aps,10pt,prd,notitlepage,nofootinbib,superscriptaddress]{revtex4-1}

\usepackage[utf8]{inputenc}
\usepackage{amsmath,amssymb,amsfonts}
\usepackage{newtxtext,newtxmath}
\usepackage{bbold}
\usepackage{bm}
\usepackage{graphicx}
\usepackage[usenames,dvipsnames]{xcolor}
\usepackage{color}
\usepackage{tikz}
\usepackage[colorlinks=true,linkcolor=blue,urlcolor=blue,citecolor=blue]{hyperref}
\usepackage{slashed}
\usepackage[english]{babel}
\usepackage{microtype}
\usepackage{dcolumn}
\usepackage{blindtext}
\usepackage{epsfig}
\usepackage{xcolor}
\usepackage{pifont}
\usepackage{dsfont,mathrsfs}
\usepackage{cancel}
\usepackage{bigints}

\usepackage{natbib}
\usepackage{accents}
\usepackage{soul}
\usepackage{multirow}
\newcommand{\nn}{\nonumber}

\newcommand{\ensembleaverage}[1]{\left\langle#1\right\rangle}
\newcommand{\MB}[1]{\left|#1\right|}
\newcommand{\FB}[1]{\left(#1\right)}
\newcommand{\SB}[1]{\left\{#1\right\}}
\newcommand{\TB}[1]{\left[#1\right]}
\newcommand{\AB}[1]{\left<#1\right>}

\newcommand{\scrL}{\mathscr{L}}

\newcommand{\munu}{{\mu\nu}}

\newcommand{\RE}{\text{Re}}

\newcommand{\fsl}{\slashed}

\newcommand{\Tr}[1]{ \text{Tr}\left[#1\right]}
\newcommand{\intzinf}{\int_{0}^{\infty}}

\newcommand{\half}{\dfrac{1}{2}}

\newcommand{\kzint}[1]{\int_{-\infty}^\infty \dfrac{d{#1}_z}{2\pi}}

\newcommand{\pbarpasi}{\AB{\bar{\psi}\psi}}

\newcommand{\diag}{\texttt{diag}}
\newcommand{\potU}{\mathcal{U}\FB{ \Phi,\bar{\Phi} ;T}}
\newcommand{\fnppbT}{\FB{\Phi, \bar{\Phi}, T } }
\newcommand{\F}{\Phi}
\newcommand{\Fb}{\bar{\Phi}}
\newcommand{\paroneder}[2]{\frac{\partial {#1}}{\partial {#2}} }
\newcommand{\partwoder}[2]{\frac{\partial^2 {#1}}{\partial{#2}^2} }
\newcommand{\parmultwoder}[3]{\frac{\partial^2 {#1}}{\partial{#2}\partial{#3}} }
\newcommand{\eBprefac}{\sum_{n,f,s} \frac{\MB{q_f B}}{2\pi^2} \intzinf dp_z  }
\newcommand{\ExpFac}[2]{e^{-{#1}\beta\FB{\omega_{nfs}  {#2}\mu_q  }} }
\newcommand{\AMMbracket}{\FB{1 - \frac{s\kappa_f q_f B}{M_{nfs}} }}

\newcommand{\replacement}{ \SB{\frac{}{} \F \leftrightarrow \Fb ; \mu_q \rightarrow -\mu_q  }}

\begin{document}
\title{Effects of quark anomalous magnetic moment on the thermodynamical properties and mesonic excitations of magnetized hot and dense matter in PNJL model}


\author{Nilanjan Chaudhuri}
\email{sovon.nilanjan@gmail.com}
\affiliation{Variable Energy Cyclotron Centre, 1/AF Bidhannagar, Kolkata - 700 064, India}
\affiliation{Homi Bhabha National Institute, Training School Complex, Anushaktinagar, Mumbai - 400085, India}
\author{Snigdha Ghosh}
\email{snigdha.physics@gmail.com, snigdha.ghosh@saha.ac.in}
\thanks{(Corresponding Author)}
\affiliation{Saha Institute of Nuclear Physics, 1/AF Bidhannagar, Kolkata - 700 064, India}
\author{Sourav Sarkar}
\email{sourav@vecc.gov.in}
\affiliation{Variable Energy Cyclotron Centre, 1/AF Bidhannagar, Kolkata - 700 064, India}
\affiliation{Homi Bhabha National Institute, Training School Complex, Anushaktinagar, Mumbai - 400085, India}
\author{Pradip Roy}
\email{pradipk.roy@saha.ac.in}
\affiliation{Saha Institute of Nuclear Physics, 1/AF Bidhannagar, Kolkata - 700 064, India}
\affiliation{Homi Bhabha National Institute, Training School Complex, Anushaktinagar, Mumbai - 400085, India}

\begin{abstract}
Various thermodynamic quantities and the phase diagram of strongly interacting hot and dense magnetized quark matter are obtained with the $ 2 $-flavour Nambu-Jona-Lasinio model with Polyakov loop considering finite values of the anomalous magnetic moment (AMM) of the quarks. Susceptibilities associated with constituent quark mass  and traced Polyakov loop are used  to evaluate chiral and deconfinement transition temperatures. It is found that, inclusion of the AMM of the quarks in presence of the background magnetic field results in a substantial decrease in the chiral as well as deconfinement transition temperatures in contrast to an enhancement in the chiral transition temperature in its absence. Using standard techniques of finite temperature field theory, the two point thermo-magnetic mesonic correlation functions in the scalar ($\sigma$) and neutral pseudoscalar ($\pi^0$) channels are evaluated to calculate the masses of  $\sigma $ and $ \pi^0 $ considering the AMM of the quarks. 
\end{abstract}
\maketitle

\section{Introduction}
	 Presence of  a finite  background magnetic field leads to a large number of exotic phenomena in strongly interacting matter. Among these some of the important ones are Chiral Magnetic Effect (CME)~\cite{Fukushima,Kharzeev,Kharzeev2,Bali}, Magnetic Catalysis (MC)~\cite{Shovkovy,Gusynin1,Gusynin2,Gusynin3}  and Inverse Magnetic Catalysis (IMC)~\cite{Preis,Preis2}  of dynamical chiral symmetry breaking which may cause significant change in the nature of electro-weak~\cite{Elmfors,Skalozub,Sadooghi_ew,Navarro}, chiral and superconducting phase transitions~\cite{Fayazbakhsh1,Fayazbakhsh2,Skokov,Fukushima2}, electromagnetically induced superconductivity and superfluidity~\cite{Chernodub1,Chernodub2} and so on. Understanding these aspects could help us to get a better picture of our main objective of understanding quantum chromodynamics (QCD). It has been reported that  strong magnetic fields of the order of $ 10^{18} $ G~\cite{Kharzeev,Skokov2} or larger may be  generated in non-central heavy-ion collisions, at RHIC and LHC which can influence substantial change in the properties of QCD matter as the magnitudes of these fields are comparable to the QCD scale i.e. $ eB\approx m_\pi^2 $ (note that in natural units, $ 10^{18} {\rm ~G} \approx m_\pi^2 \approx 0.02~ {\rm GeV}^2  $). It is conjectured that the presence of finite electrical conductivity of the hot and dense medium created during heavy ion  collisions can delay the decay of these time-dependent magnetic fields substantially~\cite{Tuchin1,Tuchin2,Gursoy}. Strong magnetic fields can be present in several other physical environments. For example, during the electroweak phase transition in the early universe  the magnetic field as high as $ \approx 10^{23}$ G~\cite{Vachaspati,Campanelli} might have been produced. At the surface and in the interior of certain compact stars called \textit{magnetars} magnetic field  of the order of $ \sim 10^{15}$ G and  $\sim 10^{18} $ G  respectively could be realized~\cite{Duncan,Duncan2,Lai}.  Moreover, observations of gravitational waves from collisions of neutron stars have triggered simulative study of such events where data for  QCD phase diagram  at large range of densities and temperatures  are required as input~\cite{Bose}. Thus study of QCD matter in these extreme conditions  has attracted a wide spectrum of researchers in this domain of physics in recent times.

	It is well known that a first principle analysis of the above mentioned phenomena is hindered due to the large coupling  strength of QCD in the low energy regime which restricts the use of perturbative approach. One of the best alternatives is to rely on Lattice QCD (LQCD) simulations. Methods, like a Taylor expansion~\cite{Bazavov2017} or an analytical continuation from imaginary chemical potentials~\cite{Gunther}, have been developed to extrapolate  thermodynamical quantities at intermediate temperatures (comparable to the QCD scale) and low baryonic density  which is relevant for highly relativistic heavy ion collisions~\cite{Bazavov2,Bazavov,Bazavov2017,Gunther,Bali_lat,Bali_lat3,Sharma}. However, for compact stars one has to consider high values of baryonic chemical potential, which are not accessible via the LQCD simulation due to the so-called sign problem in Monte Carlo sampling~\cite{Preis}.  An alternative approach is to work with effective models which are capable of incorporating most of the essential features of QCD and are mathematically tractable. Nambu-Jona-Lasinio (NJL) model~\cite{Nambu1,Nambu2} is one such model, constructed by respecting the global symmetries of QCD and it presents a useful scheme to probe arbitrary temperatures and baryonic density. This model has been extensively used  to study  the chiral symmetry restoration (see~\cite{Klevansky,Hatsuda1,Vogl,Buballa} for reviews). As mentioned in~\cite{Klevansky}, the point like interaction between quarks makes the NJL  model non-renormalizable. Thus, a proper regularization scheme is adopted to deal with the divergent integrals and the parameters associated with the model are fixed to reproduce some well known phenomenological quantities, e.g., pion-decay constant $ f_\pi  $, condensate etc~\cite{Reinder}. However, the NJL model lacks confinement:  poles of the massive quark propagator are present at any temperature and/or chemical potential. But in QCD both dynamical chiral symmetry breaking and confinement are realized as  global symmetries of the QCD Lagrangian.	It is well known that the Polyakov loop can be used as an approximate order parameter for the deconfinement transition associated with the spontaneous symmetry breaking of the center symmetry~\cite{Cheng,McLerran}. Thus, in order to obtain a unified picture of confinement	and chiral symmetry breaking the Polyakov loop enhanced Nambu-Jona-Lasinio (PNJL) model is introduced and developed by incorporating a temporal, static and homogeneous gluon-like field~\cite{Ratti,Ratti2,Ratti3,SasakipNJL,Andersen,FUKUSHIMA2004277,Fukushima_pNJLPD,FUKUSHIMA201399,Sanjay,Abhijit}. Furthermore,  the PNJL model belongs to the same universality class of QCD due to the symmetries of the Lagrangian which makes it better suited for studying the phase structure and critical phenomena related with the chiral and deconfinement phase transitions~\cite{SasakipNJL}.
	
	 PNJL model has been extensively used to study the deconfinement and chiral symmetry restoration in the presence of a background electromagnetic field~\cite{Fukushima_CME_pNJL,Gatto,Gatto2,Ferreira2,Ferreira_IMC,Mao_pNJL_GeB,Avancini_PNJL}. In~\cite{Fukushima_CME_pNJL} it is shown that the external magnetic field is likely to strengthen the chiral condensate  resulting an increase of transition temperature	compared to the zero field case in agreement with the previous studies on magnetic catalysis(MC) in NJL-like models. The modification of the phase structure of the model due to chiral chemical  potential, which mimics the chirality induced by topological excitations according to the QCD anomaly relation, has also been discussed~\cite{Fukushima_CME_pNJL}. In~\cite{Avancini_PNJL}, it has been observed that though the  electric field  partially restores the chiral symmetry, the deconfinement phase transition is marginally affected. Recent lattice results~\cite{Bali,Bali_lat,Bali_lat2,Bali_lat3} shows that although, at low temperature  the magnetic field catalyzes the chiral condensate, at higher values of  the temperature the opposite trend is observed. A combined effect of these findings indicate an overall decrease in the transition temperature leading to IMC. A significant amount of research has been conducted to explain this discrepancy by adopting appropriate modifications in the NJL-type models (see~\cite{Bandyopadhyay:2020zte,Preis2} for a review). For example, IMC is obtained in~\cite{Ahmad:2016iez,Avancini,Farias,Avancini:2019wed} by considering a lattice-inspired $ eB $-dependent coupling constant.  In~\cite{Ayala:2014gwa,Ayala:2015lta}, the effective potential was obtained beyond mean field in the linear sigma model with fermions interacting in presence of a background magnetic field and it was shown that  inclusion of  the thermo-magnetically modified couplings leads to IMC behaviour. Chiral symmetry breaking for quark matter in a magnetic background at
	 finite temperature and quark chemical potential is also studied in~\cite{ruggieri_plb}, making use of the Ginzburg–Landau effective action formalism in a renormalized quark–meson model. The observation  of IMC at finite $\mu$ up to moderate values of $eB$ is confirmed up to $eB \sim 10m_\pi^2$ in their calculations. However,  at large $eB$  magnetic catalysis is seen to appear.
	 In~\cite{Arghya}, it has been demonstrated that the inclusion of AMM of protons and neutrons leads to a decrease in critical temperature for vacuum to nuclear matter transition with increasing magnetic field which can also be identified as	IMC. Now, it is well known that quarks carry finite AMM~\cite{Sadooghi}. Thus, the main objective of our work is to include, for the first time,  the effects of the AMM of the quarks in the PNJL model and study  how the  deconfinement and chiral symmetry restoration are	 modified.A detailed study of susceptibilities related to the constituent quark mass 	 and the traced Polyakov loop are executed to evaluate the modifications in chiral and deconfinement transition temperatures due 	 to inclusion of the AMM of the quarks. Variations of quark number susceptibility, specific heat and velocity of sound are also
	 		demonstrated.

	 In addition, properties of light scalar ($ \sigma $) and pseudo-scalar ($ \pi $) mesons have also been examined in this framework to observe the effects of the Polyakov loop dynamics on the physical properties of $ \sigma,\pi $ which have a direct relevance with the dynamics of chiral symmetry restoration for hadronic systems at finite temperature and/or chemical potential. Properties of $ \sigma $ and $ \pi $ mass have already been discussed at vanishing magnetic field~\cite{Hansen,Costa_meson,Sanjay_meson,Blanquier_meson,Blaschke_meson,Costa_meson_2}. From NJL model studies~\cite{Sadooghi1,MaoPion,Avancini,Chaudhuri,Snigdha,Zhang} it is expected that the minimum temperature for	 which the overlap interval starts in the crossover region increases with the increasing magnetic field. But, we have not come across any previous calculations regarding the effects of background magnetic field or AMM of the quarks in the mesonic properties using PNJL model.  We would like to mention that all the results presented in this work have been evaluated by taking all the Landau levels of the quarks into consideration without resorting to any approximation on the strength of the magnetic field.
	
%
	The paper is organized as follows.  Sec.~\ref{Formalism} is divided in three subsections where we describe  the PNJL model very briefly (Sec.~\ref{The_model}), derivation of thermodynamic quantities (Sec.~\ref{TD_PNJL}) and mesonic properties (Sec.~\ref{Meson_prop}) respectively. Next in Sec.~\ref{sec.results} we  present the numerical results for various observables followed by a summary and conclusion of our work in Sec.~\ref{sec:conclusion}.

\section{Formalism}\label{Formalism}
\subsection{PNJL MODEL IN A HOT AND DENSE MAGNETIZED MEDIUM}\label{The_model}
The Lagrangian of the two-flavour PNJL-model considering the AMM of free quarks in presence of constant background magnetic field is given by
\begin{equation}
\scrL = \bar{\psi}(x)\FB{i\fsl{D}-m + \gamma_0 \mu_q +\half\hat{a} \sigma^\munu F_\munu  }\psi(x)  
+ G\SB{ \FB{\bar{\psi}(x) \psi (x)}^2 + \FB{ \bar{\psi}(x) i\gamma_5\tau \psi(x)}^2} - \mathcal{U} \FB{\Phi , \bar{\Phi } ; T} \label{PNJL_lagrangian}
\end{equation}
where we have dropped the flavour ($ f=u,d $) and color ($ c=r,g,b $) indices from the Dirac field $ \FB{ \psi^{fc}} $ for a convenient representation. In Eq.~\eqref{PNJL_lagrangian}, $m$ is current quark mass representing the explicit chiral  symmetry breaking (we will take $ m_u= m_d = m $ to ensure isospin symmetry of the theory at vanishing magnetic field) and  $ \mu_q $ is the chemical potential of the quark. The constituent quarks interact with the Abelian gauge field $ A_\mu  $ and the $ {\rm SU_c(3)} $
gauge field $ \mathcal{A}_\mu  $ via the covariant derivative
\begin{eqnarray}
 D_\mu = \partial_\mu  -i\hat{q}A_\mu- i \mathcal{A}_\mu^a .
\end{eqnarray}
The Abelian gauge field $ A_\mu  $ describes the influence of the external magnetic field $ B $ aligned along the $ z $-direction, for convenience we choose
$ A_\mu = \FB{0,0,xB,0} $.   The electric charges of the quarks are defined by $ \hat{q} = \diag ( 2e/3, -e/3) $\footnote{  The hat symbol on each quantity implies that they are $ 2 \times 2 $  matrices in flavor space.}. The $ {\rm SU_c(3)} $ gauge field $ \mathcal{A}_\mu  $ represents a non-trivial background due to the Polyakov loop and defined as $ \mathcal{A}_\mu =  g_s \mathcal{A}_\mu^a \lambda^a/2$  where $ g_s $ is the $ {\rm SU_c(3)} $ gauge coupling constant and $ \lambda^a $ are the Gell-Mann matrices. In the Polyakov gauge and at finite temperature $ \mathcal{A}_\mu = \delta_{\mu 0}\mathcal{A}^0 $~\cite{FUKUSHIMA2004277,Ratti,SasakipNJL}. 
In Eq.~\eqref{PNJL_lagrangian}, the factor $\hat{a}= \hat{Q}\hat{\kappa }$,  where $ \hat{\kappa}=\texttt{diag}(\kappa_u,\kappa_d) $, is a  $2\times 2 $ matrix in the flavour space. Note that, here $ \kappa_f $'s are AMM of the quarks, having dimension $ \propto 1/[M] $, defined as $ \kappa_f = \alpha_f/2M_f $ with $ M_f $ being the constituent quark mass to be defined later (note that in our case $ M_u = M_d $). $ \alpha_f $'s are dimensionless quantities defined as $ \mu_f = q_f e (1 + \alpha_f) \sigma^3/2M_f $, where $ \mu_f $ is the spin magnetic moment (see Ref.~\cite{Sadooghi} for  details).
Furthermore, $ F^\munu= \partial^\mu A^\nu- \partial^\nu A^\mu  $ and  $ \sigma^\munu=i[\gamma^\mu,\gamma^\nu]/2$. The metric tensor used in this work is  $g^\munu = \texttt{diag}\FB{1,-1,-1,-1}$. The potential $ \mathcal{U}\FB{\Phi,\bar{\Phi};T} $ in the Lagrangian (Eq.\eqref{PNJL_lagrangian}) governs the dynamics of the traced Polyakov loop and its conjugate: 
 \begin{equation}
 \Phi = \frac{1}{3} \text{Tr}_\text{c} L ~~~~~~;~~~~~~\bar{\Phi} = \frac{1}{3} \text{Tr}_\text{c} L^\dagger
 \end{equation}
 where $ L $ is the matrix in color space related to the gauge field $ \mathcal{A}_\mu $ by 
 \begin{equation}
 L(\vec{x}) = \mathscr{P} \exp\TB{i \int_0^\beta d\tau \mathcal{A}_4\FB{\vec{x},\tau  } }.
 \end{equation}
 Here $ \mathscr{P} $  denotes the  path ordering in Euclidean time, $ \beta = 1/T $ and $ \mathcal{A}_4  = i \mathcal{A}_0$. In this work we adopt the following Polyakov loop potential~\cite{Ratti} 
 \begin{eqnarray}\label{polyakov_potential}
 \frac{\mathcal{U}\FB{ \Phi,\bar{\Phi} ;T}}{T^4} = -\frac{b_2 (T)}{2}  \bar{\Phi} \Phi - \frac{b_3}{6}\FB{\Phi^3 + \bar{\Phi}^3}+ \frac{b_4}{4}\FB{ \bar{\Phi} \Phi }^2
  \end{eqnarray} 
 where 
 \begin{equation}
 b_2(T) = a_0 + a_1 \FB{\frac{T_0}{T}}+ a_2 \FB{\frac{T_0}{T}}^2+ a_3 \FB{\frac{T_0}{T}}^3.
 \end{equation}
 Values of different co-efficients are tabulated in Table-\ref{table1}~\cite{Ratti}.
 \begin{center}
 	\begin{table}[h!]
 			\caption{Parameter set for Polyakov potential}
 	\begin{tabular} { p{2cm}p{2cm}p{2cm}p{2cm}p{2cm}p{2cm}  }
 		\hline \hline
 	    $ a_0 $ & $ a_1 $ &  $ a_2 $ &  $ a_3 $  & $ b_3 $ & $ b_4 $ \\ 
 		\vspace{0.02IN} $ 6.75 $ \vspace{0.02IN}& \vspace{0.02IN}$ -1.95$ & \vspace{0.02IN} $ 2.625 $ &\vspace{0.02IN}  $ -7.44 $  &\vspace{0.02IN} $ 0.75 $ &\vspace{0.02IN} $ 7.5 $ \\
 		 \hline
 	\end{tabular}
 \label{table1}
 \end{table}
 	 \end{center}
 Following the argument in ~\cite{Ratti} we have chosen $ T_0 = 190 $ MeV.  Now expanding $ \bar{\psi } \psi  $ around the quark-condensate $ \pbarpasi $ and dropping the quadratic 
term of the fluctuation one can write
\begin{equation}
\FB{\bar{\psi} \psi }^2 = \FB{\bar{\psi} \psi -\pbarpasi +\pbarpasi}^2 \approx 2\pbarpasi \FB{\bar{\psi} \psi } - \pbarpasi^2.
\end{equation}
There is no contribution from the second term as the expectation value of the pseudo-scalar channel is zero. In this mean field approximation (MFA) and using the gauge choice for external magnetic field, the Lagrangian becomes 
 \begin{equation}
 \scrL^{MF} = \bar{\psi}(x)\FB{i\fsl{D} - M + \gamma_0 \mu_q +\hat{a} \sigma^{12}B  }\psi(x)  - { \frac{(M-m)^2}{2G}}- \potU
 \end{equation}
 where, $ M $ is the constituent quark mass given by
 \begin{equation}
 M=m - 2G\pbarpasi. \label{gap_eqn}
 \end{equation} 
  Now following   Refs. \cite{Lect_note,Fukushima_CME_pNJL}, the one-loop effective potential i.e. the thermodynamic potential for a  two-flavor Polyakov NJL model considering the AMM of the quarks at
  finite temperature ( $ T $) and chemical potential ($ \mu_q $) in presence of a uniform background magnetic field is expressed as
  \begin{eqnarray}\label{Omega_PNJL}
\Omega &=&   \frac{(M- m_0)^2}{2 G}+ \potU - 3 \sum_{n,f,s} \frac{\MB{q_f B}}{2\pi } \kzint{p}\omega_{nfs}  \nn \\&& - \frac{1}{\beta} \sum_{n,f,s} \frac{\MB{q_f B}}{2\pi } \kzint{p}\TB{\ln g^{(+)}\FB{\Phi, \bar{\Phi}, T } + \ln g^{(-)}\FB{\Phi, \bar{\Phi}, T } }
  \end{eqnarray}
  where $ \omega_{nfs} $ are the energy eigenvalues of the  quarks in the presence of external magnetic field as a consequence of the Landau quantization of the transverse momenta of the quarks and is  given by
   \begin{equation} 
   \omega_{nfs}  = \TB{p_z^2 + \SB{ \FB{\sqrt{\MB{q_f B} (2n+1-s) +M^2} - s\kappa_fq_fB }^2 }}^{\half}
   \label{energy}
   \end{equation}
   with $n$ and $ s $ being the Landau level and the spin indices respectively. The quantities $ g^{(+)}\fnppbT $ and $ g^{(-)}\fnppbT $  are defined as
   \begin{eqnarray}
  g^{(+)}\fnppbT &=& 1 + 3 \FB{  \Phi + \Fb e^ { - \beta ( \omega_{nfs} -\mu_q )} } e^ { - \beta ( \omega_{nfs} -\mu_q )} + e^ { - 3\beta ( \omega_{nfs} -\mu_q )} \\
    g^{(-)}\fnppbT &=& 1 + 3 \FB{  \Fb + \F e^ { - \beta ( \omega_{nfs} + \mu_q )} } e^ { - \beta ( \omega_{nfs} + \mu_q )} + e^ { - 3\beta ( \omega_{nfs}+ \mu_q )} .
   \end{eqnarray}
An important aspect of the PNJL model can be realized by studying  the qualitative behaviour of the thermodynamic potential at low temperature values. From Eq.~\eqref{Omega_PNJL} it is evident that in the limit $ \F,~ \Fb \rightarrow 0 $,  which is  the case at low temperatures, the contributions of one and two-quark states in the expressions of $ g^\pm  $ are strongly  suppressed compared to  the three-quark term $\sim \ExpFac{3}{\pm} $. In this sense the PNJL model mimics the confinement of quarks within three-quark states and  on a qualitative level, this is similar to the properties of QCD.  This justifies the suitability of PNJL model for describing the low-temperature QCD phase over NJL model, where the constituent quarks are abundant also at low temperatures. However, at least in the mean-field approximation, the PNJL model is deficient in one and two-quark states at low temperatures  which also  plays an important role in the investigations of the properties of  QCD. Now from Eq.~\eqref{Omega_PNJL}  one can obtain the expressions for the constituent quark mass ($ M $) and the expectation values of the Polyakov loops $ \F $ and $ \Fb $ using the following stationary conditions: 
\begin{equation}
\paroneder{\Omega}{M} = 0 ; ~~~~~~~~ \paroneder{\Omega}{\F} = 0 ;  ~~~~~~~~ \paroneder{\Omega}{\Fb} =0;
\end{equation} 
which leads to the following sets of coupled integral equations
\begin{eqnarray}
&& M = m + 3G \sum_{n,f,s} \frac{\MB{q_f B }}{2\pi^2 } \intzinf dp_z \frac{M}{\omega_{nfs}} \FB{1 -  \frac{s\kappa_f q_f B}{M_{nfs}}}\nn \\ && -3G \sum_{n,f,s} \frac{\MB{q_f B}}{2\pi^2 } \intzinf dp_z \frac{M}{\omega_{nfs} } \FB{ 1 - \frac{ s\kappa_f q_f B}{M_{nfs}}  } \TB{ \frac{}{}  f^+ \fnppbT + f^- \fnppbT} \label{Gap_M}~, \\
&& \SB{- \frac{b_2 ( T )}{2}\Fb - \frac{b_3}{2}\F^2 + \frac{b_4}{2}\FB{ \Fb \F  }\F     } - \frac{3}{T^3} \sum_{n,f,s} \frac{\MB{q_f B}}{2 \pi^2 } \intzinf dp_z \TB{\frac{e^{- \beta( \omega_{nfs} -\mu_q) }}{g^{(+)}}  + \frac{e^{- 2\beta( \omega_{nfs} +\mu_q) }}{g^{(-)}}        } = 0 \label{Gap_p}~, \\
&& \SB{- \frac{b_2 ( T )}{2}\F - \frac{b_3}{2}\Fb^2 + \frac{b_4}{2}\FB{ \Fb \F  }\Fb     } - \frac{3}{T^3} \sum_{n,f,s} \frac{\MB{q_f B}}{2 \pi^2 } \intzinf dp_z \TB{\frac{e^{- 2\beta( \omega_{nfs} -\mu_q) }}{g^{(+)}}  + \frac{e^{- \beta( \omega_{nfs} +\mu_q) }}{g^{(-)}}        } = 0 \label{Gap_pb}
\end{eqnarray}
where 
\begin{eqnarray}
M_{nfs} &=& \sqrt{ \MB{q_f B} \FB{2n + 1 -s   }  + M^2  }~, \\
f^+ \fnppbT &=& \frac{	\FB{\F + 2\Fb e^{-\beta( \omega_{nfs}  -\mu_q ) }} e^{-\beta( \omega_{nfs}  -\mu_q ) } + e^{-3\beta( \omega_{nfs}  -\mu_q ) }  	}{ 1 + 3 \FB{  \Phi + \Fb e^ { - \beta ( \omega_{nfs} -\mu_q )} } e^ { - \beta ( \omega_{nfs} -\mu_q )} + e^ { - 3\beta ( \omega_{nfs} -\mu_q )}  }~,\label{eq.fplus} \\
f^- \fnppbT &=& \frac{	\FB{\Fb + 2\F e^{-\beta( \omega_{nfs}  +\mu_q ) }} e^{-\beta( \omega_{nfs}  +\mu_q ) } + e^{-3\beta( \omega_{nfs} + \mu_q ) }  	}{ 1 + 3 \FB{  \Fb + \F e^ { - \beta ( \omega_{nfs} +\mu_q )} } e^ { - \beta ( \omega_{nfs} +\mu_q )} + e^ { - 3\beta ( \omega_{nfs} +\mu_q )}  }~. \label{eq.fminus} 
\end{eqnarray}
Note that in Eq.~\eqref{Gap_M}, the medium independent integral is ultraviolet divergent. Since the theory is known to be non-renormalizable owing to the point-like interaction between the quarks, a proper regularization scheme is necessary. Regularization schemes to handle such divergences are discussed in~\cite{Chaudhuri,Morimoto:2018pzk,Avancini_reg}.

\subsection{THERMODYNAMIC QUANTITIES}\label{TD_PNJL}
 The thermodynamics of the PNJL model in presence of the background magnetic field can be characterized by the potential $ \Omega $ defined in Eq.~\eqref{Omega_PNJL}. Since the system is uniform, pressure and energy density are given by~\cite{Buballa}
 \begin{eqnarray}
 p(T,\mu_q) &=& - \Omega( M, \F, \Fb, T, \mu_q )~,	\\
 \varepsilon(T,\mu_q) &=& -p(T,\mu_q) + T s(T,\mu_q) + \mu_q n_q(T,\mu_q )
 \end{eqnarray}
 where, $ n_q $ is the quark number density given by
 \begin{equation}
 n_q(T,\mu_q) = - \paroneder{\Omega}{\mu_q} =  3 \sum_{n,f,s}  \frac{\MB{q_f B}}{2\pi^2 } \intzinf dp_z \FB { f^+\fnppbT - f^-\fnppbT   }
   \end{equation}
while the entropy density ($ s(T, \mu_q)  $) is  defined as
\begin{eqnarray}
	s(T,\mu_q )&& = -\paroneder{\Omega}{T } = \eBprefac \TB{\ln g^{(+)}\fnppbT + \ln g^{(-)}\fnppbT   } + 3 T \eBprefac \TB{ \frac{\omega_{nfs} -\mu_q  }{T^2} f^+\fnppbT   \nn \right. \\ &&\left.  + \frac{\omega_{nfs} +\mu_q  }{T^2} f^-\fnppbT  } - 4T^3 \TB{ -\frac{b_2(T)}{2}\Fb \F - \frac{b_3}{6}\FB{\F^3 + \Fb^3 } + \frac{b_4}{4} \FB{\Fb\F}^2 } -\half \TB{   a_1 T_0 T^2 + 2  a_2 T_0^2 T+ 3a_3 T_0^3} \Fb \F .
\end{eqnarray}
During the derivation of the expression for entropy, we have used the gap equations for $ M, \F $ and $ \Fb $ given in Eqs.~\eqref{Gap_M}, \eqref{Gap_p} and \eqref{Gap_pb} respectively to get rid of the term involving $ T $-derivatives of  $ M, \F $ and $ \Fb $. The response of $ n_q $ and $ s $ due to the variations of $ \mu_q $ and $ T $ can be measured by the quark number susceptibility ($ \chi_q $) and the specific heat ($ C_V $) respectively. They can be defined as
\begin{eqnarray}
 \chi_q = \paroneder{n_q}{\mu_q} &=& \chi^0_{q} + T^2 A_{M,\mu_q} \FB{\paroneder{M}{\mu_q}}+ T^3 A_{\F ,\mu_q} \FB{\paroneder{\F}{\mu_q}}+ T^3 A_{\Fb,\mu_q} \FB{\paroneder{\Fb }{\mu_q}} \label{qnumsus} \\
 C_V  = T\FB{\paroneder{s}{T}}_V &=& T C_V^0 +  T^3~ A_{M,T} \FB{\paroneder{M}{T}} ~+ T^4 ~A_{\F ,T} \FB{\paroneder{\F}{T}} ~+ T^4~ A_{\Fb,T} \FB{\paroneder{\Fb }{T}} \label{cv} 
\end{eqnarray}
where
\begin{eqnarray}
\chi^0_q &=& \frac{3}{T} \eBprefac \TB{ \frac{\ExpFac{}{-} }{{g^{(+)}}^2}  \SB{  \F + 4\Fb \ExpFac{ }{-} + 3\FB{ 1 + \Fb \F  }\ExpFac{2}{-} + 4\F \ExpFac{3}{-} \nn \right.\right. \\ && \left.\left. + \Fb \ExpFac{4}{-}  }   \frac{}{} + \SB{\F \leftrightarrow \Fb ; \mu_q \rightarrow -\mu_q }  } \label{Chi0} 
  \end{eqnarray}
  and 
  \begin{eqnarray}
  C_V^0 &=& \frac{3}{T^3} \eBprefac \TB{ \frac{\ExpFac{}{-} }{{g^{(+)}}^2} \FB{\omega_{nfs} -\mu_q }^2 \SB{  \F  + 4\Fb \ExpFac{ }{-} + 3\FB{ 1 + \Fb \F  }\ExpFac{2}{-}  \nn \right.\right. \\ && \left.\left.+4\F \ExpFac{3}{-}  + \Fb \ExpFac{4}{-}  }   \frac{}{} + \SB{\F \leftrightarrow \Fb ; \mu_q \rightarrow -\mu_q }  } - 12T^2 \SB{ -\frac{b_2(T)}{2}\Fb \F - \frac{b_3}{6}\FB{\F^3 + \Fb^3 } + \frac{b_4}{4} \FB{\Fb\F}^2 }   \nn \\ &&  + 2T^3 \paroneder{b_2(T)}{T} \Fb \F  - \FB{a_1 T_0 T + a_2 T_0^2 } \Fb \F~.
  \label{cv0}
  \end{eqnarray}
 All the other terms appearing in Eqs.~\eqref{qnumsus} and~\eqref{cv0} are defined in Appendices   \ref{T_der} and \ref{mu_der}. One can also calculate the velocity of sound ($ c_s $) which is closely related to $ C_V $ and is given by
\begin{equation}\label{velsound}
c_s^2 = \FB{\paroneder{p}{\varepsilon}} = \frac{s}{C_V}.
\end{equation}
 Now as discussed in~\cite{SasakipNJL}, the constituent quark mass and the Polyakov loops are effective fields associated with the order parameters of  chiral and  $ Z(3) $ symmetry. Hence the susceptibilities  corresponding to these fields show signals of phase transitions. In order to calculate them we introduce the following dimensionless matrix
\begin{equation}\label{mat_C}
\mathbf{C}=\begin{bmatrix}
C_{MM}   & C_{M\F} & C_{M\Fb}  \\ C_{\F M}   & C_{\F \F } & C_{\F\Fb}\\ C_{\Fb M}   & C_{\Fb \F } & C_{\Fb \Fb}  
\end{bmatrix} 
\end{equation}
with
\begin{eqnarray}
&&C_{MM} = \frac{1}{T\Lambda} \partwoder{\Omega}{M}; ~~~~~~~~ C_{M\F} = \frac{1}{T\Lambda^2}\parmultwoder{\Omega}{M}{\F} = C_{\F M}; ~~~~~~~~ C_{M\Fb} = \frac{1}{T\Lambda^2}\parmultwoder{\Omega}{M}{\Fb} = C_{\Fb M}; \nn  \\ &&~~
C_{\F \F} = \frac{1}{T\Lambda^3} \partwoder{\Omega}{\F}; ~~~~~~~~ C_{\F \Fb } = \frac{1}{T\Lambda^3}\parmultwoder{\Omega}{\F}{\Fb} = C_{\Fb \F}; ~~~~~~~~~~~ C_{\Fb \Fb} = \frac{1}{T\Lambda^3} \partwoder{\Omega}{\Fb}.
 \end{eqnarray}
In Appendix~\ref{double_der} we have calculated different double derivatives of $ \Omega $ with respect to $ M,\F,\Fb $. Susceptibilities are defined as the inverse of $ \mathbf{C} $ and can be expressed as
\begin{equation}
\mathbf{\chi} = \mathbf{C}^{-1} =
 \begin{bmatrix}
 \chi_{MM}   & \chi_{M\F} & \chi_{M\Fb}  \\ \chi_{\F M}   & \chi_{\F \F } & \chi_{\F\Fb}\\ \chi_{\Fb M}   & \chi_{\Fb \F } & \chi_{\Fb \Fb} 
\end{bmatrix}
\end{equation}
Here $ \chi_{MM} , \chi_{\F\F}$ and $ \chi_{\Fb\Fb} $ are chiral and diagonal Polyakov loop susceptibilities respectively. The off-diagonal terms are mixed susceptibilities. Note that one can find out the $\kappa_f \rightarrow 0  $ and  $ eB\rightarrow 0 $ limit of the results obtained in this section, Sec.~\ref{The_model} and the Appendices by making the following replacements:
\begin{eqnarray}
\eBprefac &\longrightarrow& 2N_f \int \frac{d^3 \vec{p}}{\FB{2\pi}^3}~, \\
\omega_{nfs}  &\longrightarrow& E_{\vec{p} } = \sqrt{\vec{p}^2 + M^2  }~, \\
\frac{M}{\omega_{nfs}}\AMMbracket &\longrightarrow& \frac{M}{E_{\vec{p}}}
\end{eqnarray}
and 
\begin{eqnarray}
\frac{1}{\omega_{nfs}}\AMMbracket - \frac{M^2}{\omega_{nfs}^3} \AMMbracket^2 + \frac{M^2 s\kappa_f q_f B}{\omega_{nfs} M_{nfs}^3} \longrightarrow  
\FB{\frac{1}{E_{\vec{p}}} - \frac{M^2}{E_{\vec{p}}^3}}.
\end{eqnarray}
A discussion on these replacements and analytical derivation of the limiting procedure can be found in~\cite{somnath,Menenzes,Chaudhuri}.


\subsection{MESONIC PROPERTIES}\label{Meson_prop}
The mesons being the bound states of quarks and anti-quarks, their propagations can be studied within the PNJL model using the 
Bethe-Salpeter equation~\cite{Klevansky}. We are interested in the evaluation of two-point mesonic correlation functions of the type:
\begin{eqnarray}
C_a(q) = i\int d^4x e^{iq\cdot x} \ensembleaverage{\mathcal{T}J_a(x)J_a^\dagger(0)}
\label{eq.correlator-1}
\end{eqnarray}
where $\mathcal{T}$ is the time ordering symbol and $J_a(x)$ represents the  local current for the channel $a\in\{\pi^0,\sigma\}$ given by
\begin{eqnarray}
J_{\pi^0}(x) &=& \bar{\psi }(x)i\gamma^5\tau^3\psi(x) ~~~~~ (\text{pseudoscalar}),\\
J_\sigma(x) &=& \bar{\psi}(x)\psi(x) ~~~~~ (\text{scalar})
\end{eqnarray}
with $\tau^3$ being the third component of Pauli matrices in isospin space. In the Random Phase Approximation (RPA), the correlator 
in Eq.~\eqref{eq.correlator-1} can be recast into the form of a Dyson-Schwinger equation~\cite{Hansen}in the following way:
\begin{eqnarray}
C_a(q) = \Pi_a(q) + \Pi_a(q)(2G) C_a(q)
\end{eqnarray}
where, $\Pi_a(q)$ is the one-loop in-medium polarization function of the mesons. Its explicit form is given by~\cite{Hansen,Chaudhuri}
\begin{eqnarray}
\Pi_a(q) = i\int\frac{d^4k}{(2\pi)^4}\text{Tr}_\text{d,f,c}\TB{S(k)\Gamma_aS(p=q+k)\Gamma_a} ~~;~~ a\in\{\pi^0,\sigma\}.
\end{eqnarray}
Here,  $S(k)$ is the dressed Hartree quark propagator and $\text{Tr}_\text{d,f,c}$ represents the trace over the Dirac, colour 
and flavour spaces. In the above equation $\Gamma_{\pi^0}=i\gamma^5\tau^3$ and $\Gamma_\sigma=1$. The polarization functions of 
 $\pi^0$ and $\sigma$ mesons in the NJL model are explicitly calculated in Ref.~\cite{Chaudhuri} employing thermal field 
theoretic methods at both  vanishing as well as non-vanishing external magnetic field. In Ref.~\cite{Hansen}, the mesonic polarization 
functions are calculated at $B=0$ within both  NJL and PNJL models where it has been demonstrated that, going from NJL to PNJL 
model requires only the replacement of the Fermi-Dirac distribution functions of the quarks and antiquarks with the functions 
given in Eqs.~\eqref{eq.fplus} and \eqref{eq.fminus} respectively. Therefore, following Refs.~\cite{Chaudhuri,Hansen}, 
the thermal polarization functions in the PNJL model at $B=0$ and at vanishing three momentum of the mesons can be written as
\begin{eqnarray}
\RE{\Pi}_a (q^0,\vec{q}=\vec{0},B=0) &=&  \dfrac{1}{4\pi^2}\int_0^\Lambda \vec{k}^2 d|\vec{k}| 
\FB{\dfrac{1}{\omega_k q_0} } \mathcal{P} \TB{ \dfrac{\mathcal{N}_a(k^0= -q^0 +\omega_k)}{q^0-2\omega_k} 
	+ \dfrac{\mathcal{N}_a(k^0= \omega_k)}{q^0+2\omega_k}} \nn \\ 
&& - \dfrac{1}{4\pi^2} \intzinf \vec{k}^2 d|\vec{k}|  \FB{\dfrac{1}{\omega_k q_0}   } \mathcal{P} 
\TB{ \dfrac{\mathcal{N}_a(k^0= -\omega_k) f^-(\omega_k)}{q^0-2\omega_k} 
	+ \dfrac{\mathcal{N}_a(k^0= \omega_k)f^+(\omega_k) }{q^0+2\omega_k}  \nn \right. \\ 
	&& \hspace{2cm} \left. +  \dfrac{\mathcal{N}_a(k^0= -q^0 -\omega_k) f^-(\omega_k) }{q^0-2\omega_k} 
	+ \dfrac{\mathcal{N}_a(k^0= -q^0+\omega_k)f^+(\omega_k)}{q^0+2\omega_k}}
\label{eq.pola_0} 
\end{eqnarray} 
where the Cauchy principal value integral is denoted by $\mathcal{P}$ and $\mathcal{N}_a(k,q)$'s (for $ a=\sigma, \pi_0 $) are given by 
\begin{eqnarray}
\mathcal{N}_\sigma(k,q) &=& 3N_f\Tr{(\cancel{k}+\cancel{q}+M)(\cancel{k}+M)} = 12N_f(M^2 +k^2 +k\cdot q ), \\
\mathcal{N}_{\pi^0} (k,q) &=& -3N_f\Tr{\gamma^5(\cancel{k}+\cancel{q}+M)\gamma^5(\cancel{k}+M)} = -12N_f(M^2 -k^2 -k\cdot q )~. 
\end{eqnarray}
On the other hand, we have the following expressions for the thermo-magnetic polarization function in the PNJL model at  $ \vec{q} = 0 $:
\begin{eqnarray}
\RE{\Pi}_a (q^0,\vec{q}=\vec{0},B\ne0) &=& \sum_f \sum_{s_k,s_p} \sum_{l=0}^{\infty} \TB{\textcolor{white}{\dfrac{\dfrac{}{}}{\dfrac{}{}}}
\int_{0}^{\sqrt{\Lambda^2-\vec{k}_{\perp l}^2}} \dfrac{dk_z}{\pi}  
\Theta\FB{\vec{k}_{\perp l}^2} \Theta\FB{\vec{p}_{\perp l}^2} \Theta\FB{\Lambda^2-\vec{k}_{\perp l}^2} \Theta\FB{\Lambda^2-\vec{p}_{\perp l}^2} \nn\right. \\ 
&& \left. \times ~\mathcal{P} \SB{\dfrac{\mathcal{N}^a_{ls_ks_p}\FB{k^0=-q^0+\omega^{ls_p}_{k}}}{2 \omega^{ls_p}_{k} 
		\SB{\FB{q^0 - \omega^{ls_p}_{k}}^2 - \FB{\omega^{ls_k}_{k}}^2 }} 
	+\dfrac{\mathcal{N}^a_{ls_ks_p}(k^0=\omega^{ls_k}_{k})}{2 \omega^{ls_k}_{k} \SB{\FB{q^0 + \omega^{ls_k}_{k}}^2 - \FB{\omega^{ls_p}_{k}}^2 }} } \nn\right. \\ && \left.
+ \int_{-\infty}^{+\infty} \dfrac{dk_z}{(2\pi)}  
\Theta\FB{\vec{k}_{\perp l}^2} \Theta\FB{\vec{p}_{\perp l}^2}
\mathcal{P} \SB{
	- \dfrac{\mathcal{N}^a_{ls_ks_p}(k^0=-\omega^{ls_k}_{k}) f^-(\omega^{ls_k}_{k}) }{2 \omega^{ls_k}_{k} 
		\SB{\FB{q^0 - \omega^{ls_k}_{k}}^2 - \FB{\omega^{ls_p}_{k}}^2  }  } - \dfrac{\mathcal{N}^a_{ls_ks_p}(k^0=\omega^{ls_k}_{k}) f^+(\omega^{ls_k}_{k})}
	{2 \omega^{ls_k}_{k} \SB{\FB{q^0 + \omega^{ls_k}_{k}}^2 - \FB{\omega^{ls_p}_{k}}^2  }  }   \right. \nn \right. \\  && \left. \hspace{0cm} \left.
	-\dfrac{\mathcal{N}^a_{ls_ks_p}(k^0=-q^0-\omega^{ls_p}_{k}) f^-(\omega^{ls_p}_{k})   }{2 \omega^{ls_p}_{k} 
		\SB{\FB{q^0 + \omega^{ls_p}_{k}}^2 - \FB{\omega^{ls_k}_{k}}^2  }  } - \dfrac{\mathcal{N}^a_{ls_ks_p}(k^0=-q^0 +\omega^{ls_p}_{k}) f^+(\omega^{ls_p}_{k})}
	{2 \omega^{ls_p}_{k} \SB{\FB{q^0 - \omega^{ls_p}_{k}}^2 - \FB{\omega^{ls_k}_{k}}^2  }  } 
}}
\label{eq.pola.1}
\end{eqnarray}
where the flavour index $f$ in different terms within the square bracket of the right hand side of the above equation has been suppressed and 
\begin{eqnarray}
\omega^{ls_k}_k &=& \sqrt{k_z^2+(M_l-s_k\kappa eB)^2}~, \\
\vec{k}_{\perp l}^2 &=& 2 leB +\FB{\kappa e B}^2 - 2s_kM_{l}(\kappa eB)~,  \\
\vec{p}_{\perp l}^2 &=& 2 leB +\FB{\kappa e B}^2 - 2s_pM_{l}(\kappa eB)~
\end{eqnarray}
with $ M_l = \sqrt{M^2 + 2l\MB{eB}} $. The expression for $\mathcal{N}^a_{ls_ks_p}$ is given by
\begin{eqnarray}
\mathcal{N}^{a}_{ls_ks_p}(k,q) &=& 6j_a \frac{eB}{4\pi M_l^2} 
\FB{1-\delta^0_l \delta^{-1}_{s_k }}\FB{1-\delta^0_l \delta^{-1}_{s_p }} 
\Bigg[ -4eBl \Big\{ s_ks_p(k_\parallel^2+k^0q^0)
+j_as_ks_p(\kappa eB)^2 +j_aM_l^2  -j_a \kappa eBM_l(s_p+s_k) \Big\} \nn \\ &&
+ \FB{1-\delta_l^0} \Big\{j(k_\parallel^2+k^0q^0)
(M_l-s_kM)(M_l-s_pM) + \big\{s_k\kappa eBM-M_l(M-s_kM_l+\kappa eB)\big\}\big\{s_p\kappa eBM \nn \\ &&
-M_l(M-s_pM_l+\kappa eB)\big\}\Big\} + j_a(k_\parallel^2+k^0q^0)
(M_l+s_kM)(M_l+s_pM) \nn \\ && + \big\{s_k\kappa eBM-M_l(M+s_kM_l-\kappa eB)\big\}\big\{s_p\kappa eBM-M_l(M+s_pM_l-\kappa eB)\big\}
\Bigg] 
\end{eqnarray}
with $j_{\sigma}=1$ and $j_{\pi^0}=-1$. The different step functions appearing on the rhs of Eq.~\eqref{eq.pola.1} represent 
the UV and AMM blocking as discussed in~\cite{Chaudhuri}.

Having obtained the polarization functions of the mesons, it is now straightforward to evaluate the 
masses of $\pi^0$ and $\sigma$ by solving the following transcendental equations
\begin{eqnarray}
1 - 2G \Pi_a(q^0=m_a,\vec{q}=\vec{0}) = 0~, ~~~~~~~~~~ a \in \SB{\sigma,\pi^0}
\label{eq.meson.mass}
\end{eqnarray}
representing the pole of the meson propagators.


\section{Numerical Results} \label{sec.results}
In this section, we present numerical results for the dynamically generated constituent quark mass ($ M $), expectation values of Polyakov loops $ \F $ and $ \Fb $ as well as several thermodynamic quantities in a hot and dense magnetized medium considering finite values of the AMM of the quarks.   Following Refs.~\cite{Ratti, Ratti2}, we have chosen the three momentum cutoff $ \Lambda=651 $ MeV, coupling constant $ G=10.08~{\rm GeV^{-2}} $ and bare quark mass $ m=5.5$ MeV.  These parameters have been fixed by fitting the empirical values of pion mass $ m_\pi=139.3 $ MeV and pion decay constant $ f_\pi=92.4$ MeV at zero temperature and zero baryon density in the absence of the background magnetic field. For these values of parameters we obtain $ \MB{\pbarpasi}^{1/3} = 251 $ MeV and $ M=325$ MeV at $ T\rightarrow 0, \mu_q \rightarrow 0 $.  We have considered constant values of AMM of the quarks, $ \kappa_u= 0.29 \ {\rm GeV^{-1}}$  and $ \kappa_d= 0.36 \ {\rm GeV^{-1}}$ following Ref.~\cite{Sadooghi}.

\begin{figure}[h]
	\begin{center}
		\includegraphics[angle=-90, scale=0.35]{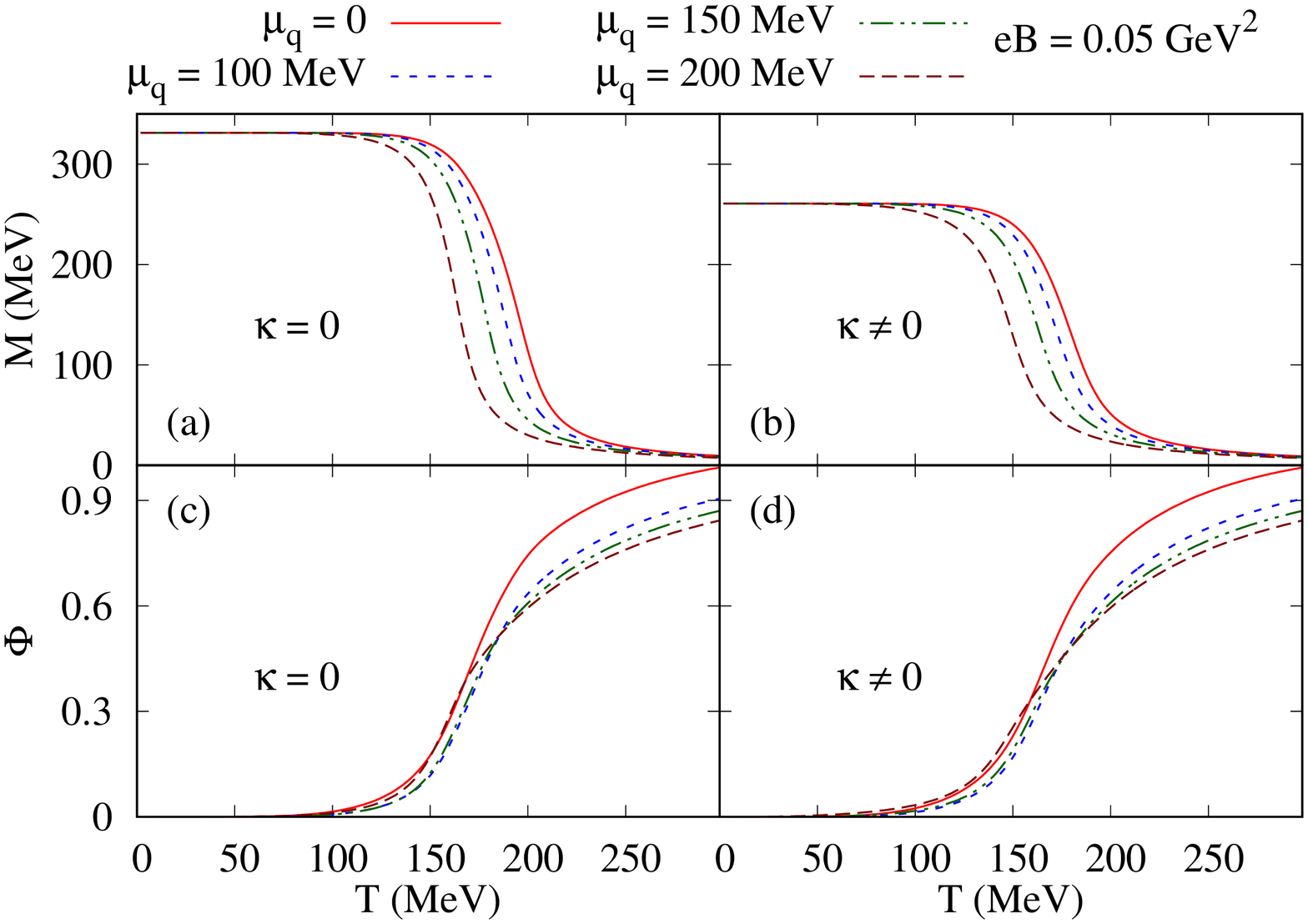}					
		\includegraphics[angle=-90, scale=0.35]{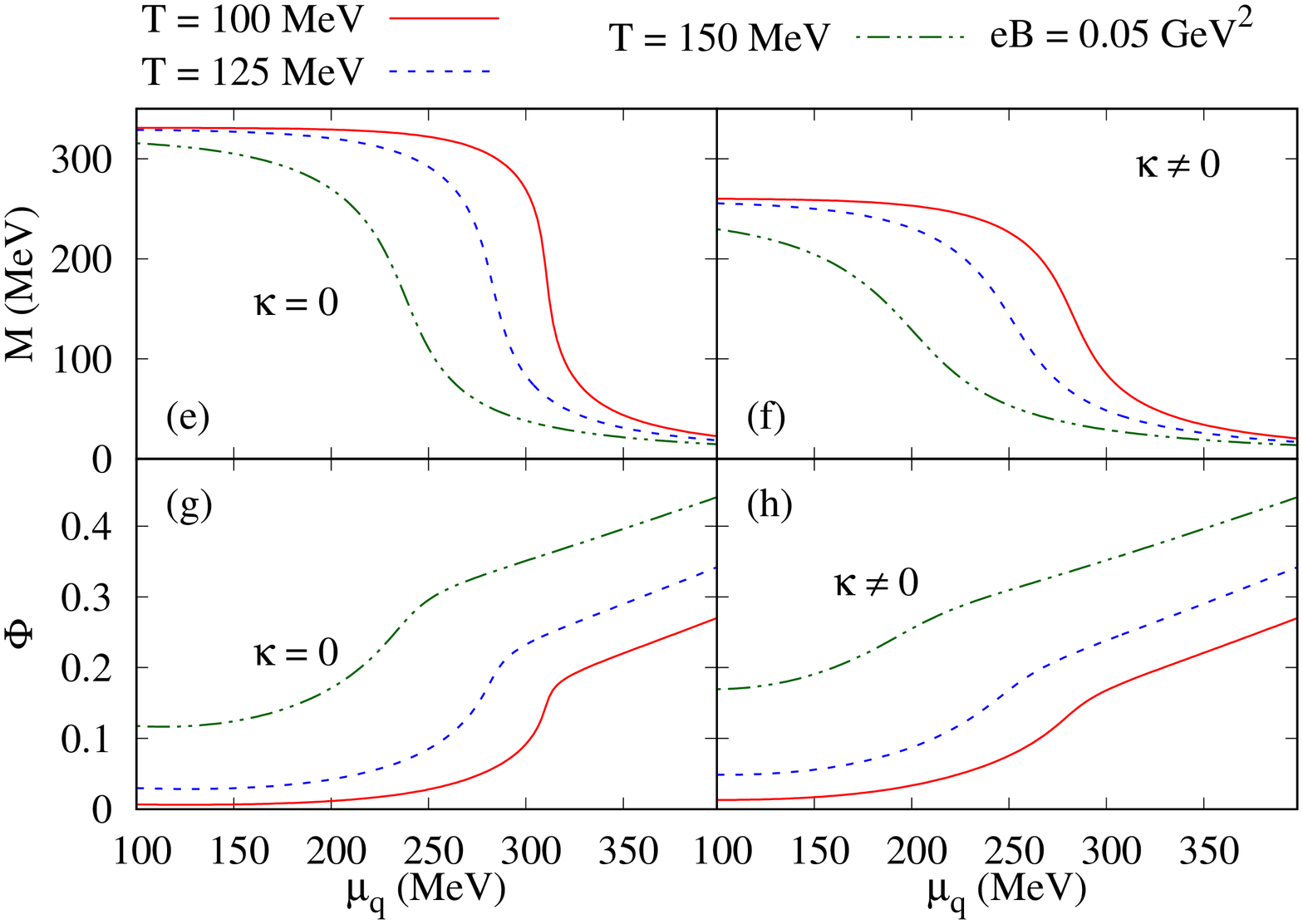}					
	\end{center}
	\caption{ Variation of $ M $ and $ \F  $ as a function of $ T $ and $ \mu_q $ at $ eB= 0.05 {\rm ~GeV^2} $ with and without considering the AMM of the quarks. }
	\label{MF_vs_Tmu}
\end{figure}
\begin{figure}[h]
	\begin{center}
		\includegraphics[angle=-90, scale=0.35]{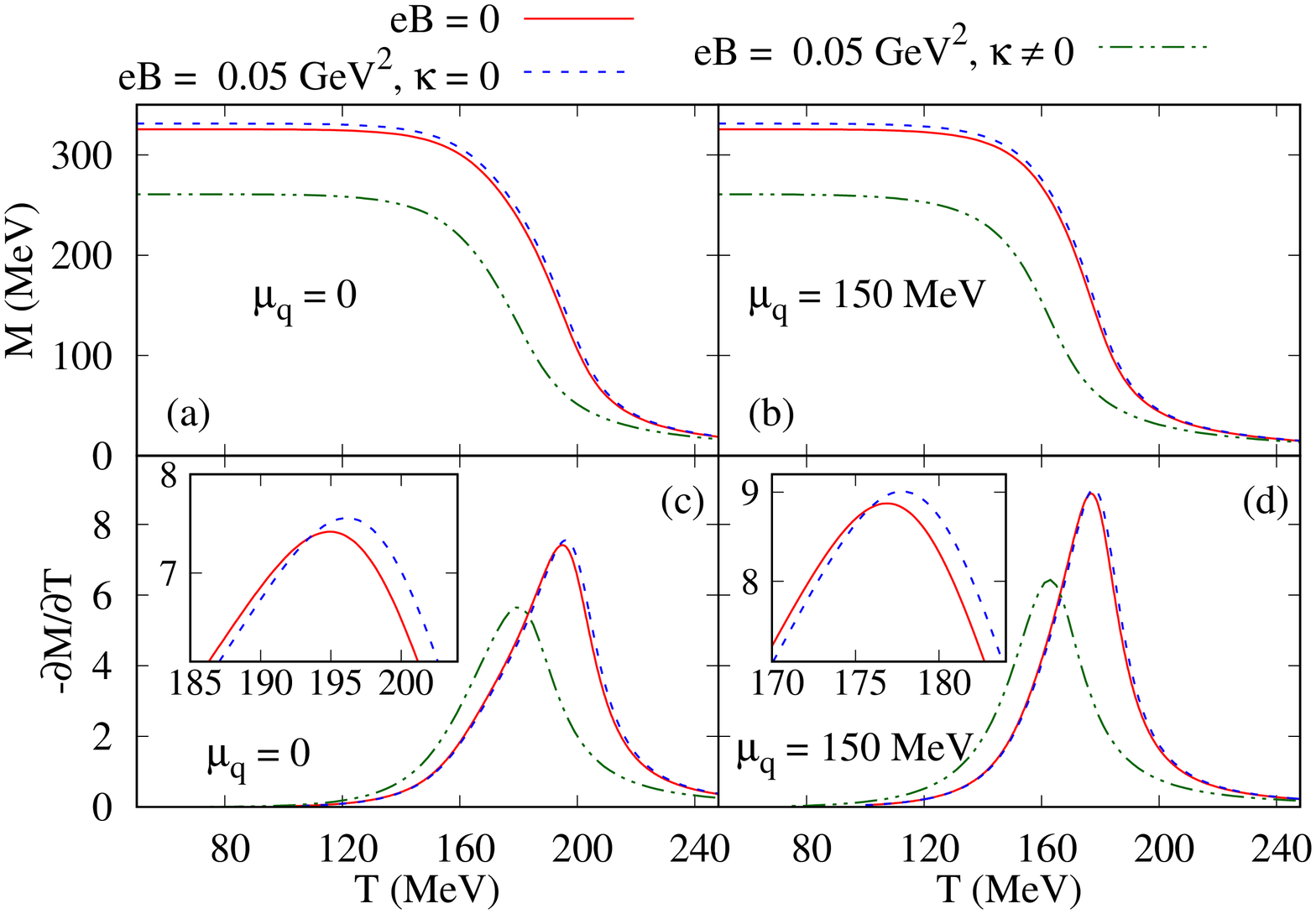}					
		\includegraphics[angle=-90, scale=0.35]{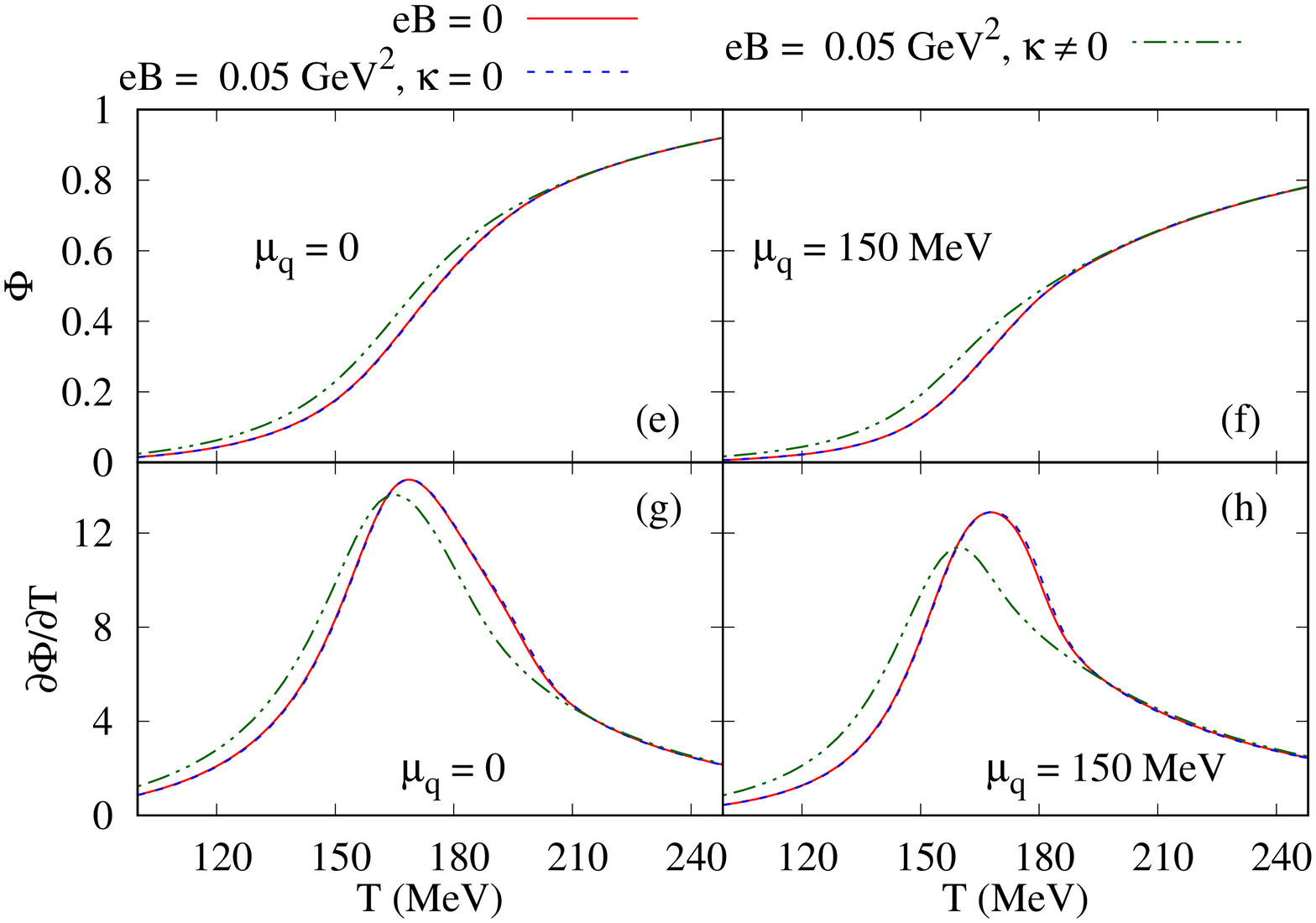}					
	\end{center}
	\caption{ Variation of $ M $, $ -\FB{\paroneder{M}{T}} $, $ \F $ and $ \FB{\paroneder{\F}{T}} $ as a function of $ T $ at different values of $ \mu_q $ and $ eB $  considering zero and non-zero values of the AMM of the quarks. Inset plots in (c) and (d) show the two peaks by enlarging the relevant temperature region.}
	\label{MF0eBk_vs_T}
\end{figure}

In Figs.\ref{MF_vs_Tmu} (a) and (b) we have shown the variation of constituent quark mass ($ M $) as a function of temperature ($ T  $) for zero and non-zero values of AMM of the quarks in the presence of a uniform background magnetic field i.e. at $ eB = 0.05 ~{\rm GeV^2} $. In both the plots we have varied the chemical potential as $ \mu_q = 0, 100, 150 ~{\rm and}~ 200 {\rm ~ MeV} $. Comparing Figs.\ref{MF_vs_Tmu} (a) and (b), it can be seen that there are two immediate effects of the consideration of AMM of the quarks. Firstly, it leads to significant decrease of $ M $ in the limit $ T\rightarrow 0 $. Secondly, the transition from chiral symmetry broken to the restored phase occurs at lower values of temperature for all $ \mu_q $ values. We will come back to this behaviour of $ M $ later while describing Figs.~\ref{MF0eBk_vs_T} (a)-(d).  The overall behaviour of $ M $ is qualitatively similar in both the cases as it starts from a high value at low $ T $, remains almost constant up to $ T \approx 100 ~{\rm MeV}  $ and finally becomes nearly equal to the bare quark mass.   Thus, the transition from the chiral symmetry broken to the restored phase is a crossover. Note that, since we have considered finite value of the bare quark mass i.e. $ m_0  = 5.5 ~{\rm MeV}$, the chiral symmetry is never restored fully. However, as we increase $ \mu_q $, the crossover pattern moves towards lower values of $ T $ in both occasions. 

In Figs.~\ref{MF_vs_Tmu} (c) and (d) the expectation  value of the Polyakov loop ($ \F $) is plotted as a function of $ T $  for vanishing and non-vanishing  values of AMM of the quarks respectively at constant background magnetic field $ eB = 0.05 ~{\rm  GeV^2} $ for different values of the quark chemical potential ($ \mu_q = 0, 100, 150 ~{\rm and}~ 200 ~{\rm MeV} $) . As described in~\cite{SasakipNJL},  although the Polyakov potential introduced in Eq.~\eqref{polyakov_potential}  is $ Z(3) $ symmetric, due to the interaction with quarks this symmetry is explicitly broken. Thus, the transition from confined to deconfined phase is a rapid crossover in all the cases considered in the  above mentioned plots. However, as we include finite $ \mu_q $, Polyakov loop $ \F $ keeps on decreasing with increasing values of $ \mu_q $.    It is interesting to note that, AMM of the quarks affects the temperature variation of $ \F $ marginally. This is opposite when compared to constituent quark mass,  as we have already seen a significant decrease in $ M $ due to the consideration of finite AMM of the quarks (see Figs.~\ref{MF_vs_Tmu} (a) and (b)). 

\begin{figure}[h]
	\begin{center}
		\includegraphics[angle=-90, scale=0.35]{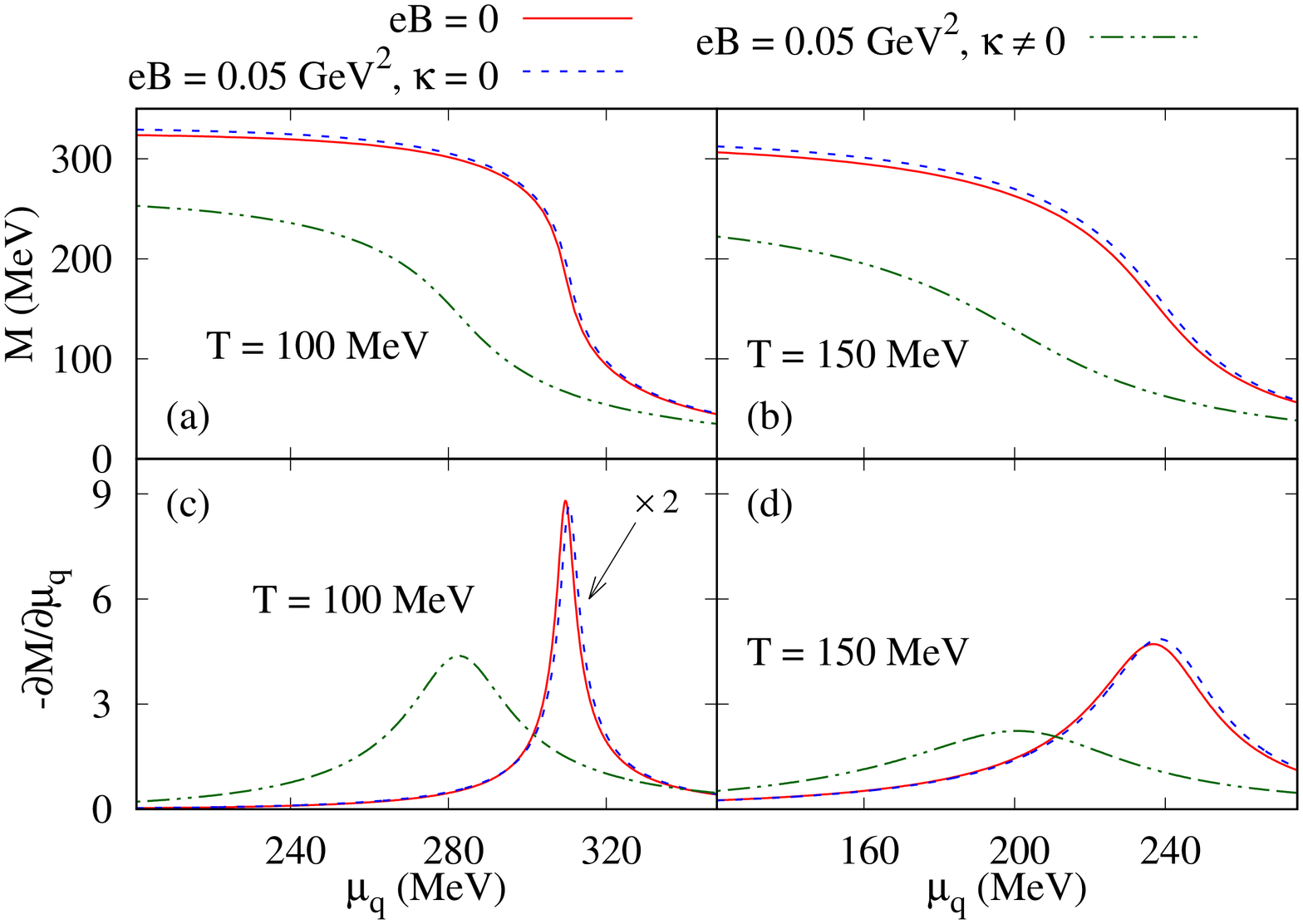}					
		\includegraphics[angle=-90, scale=0.35]{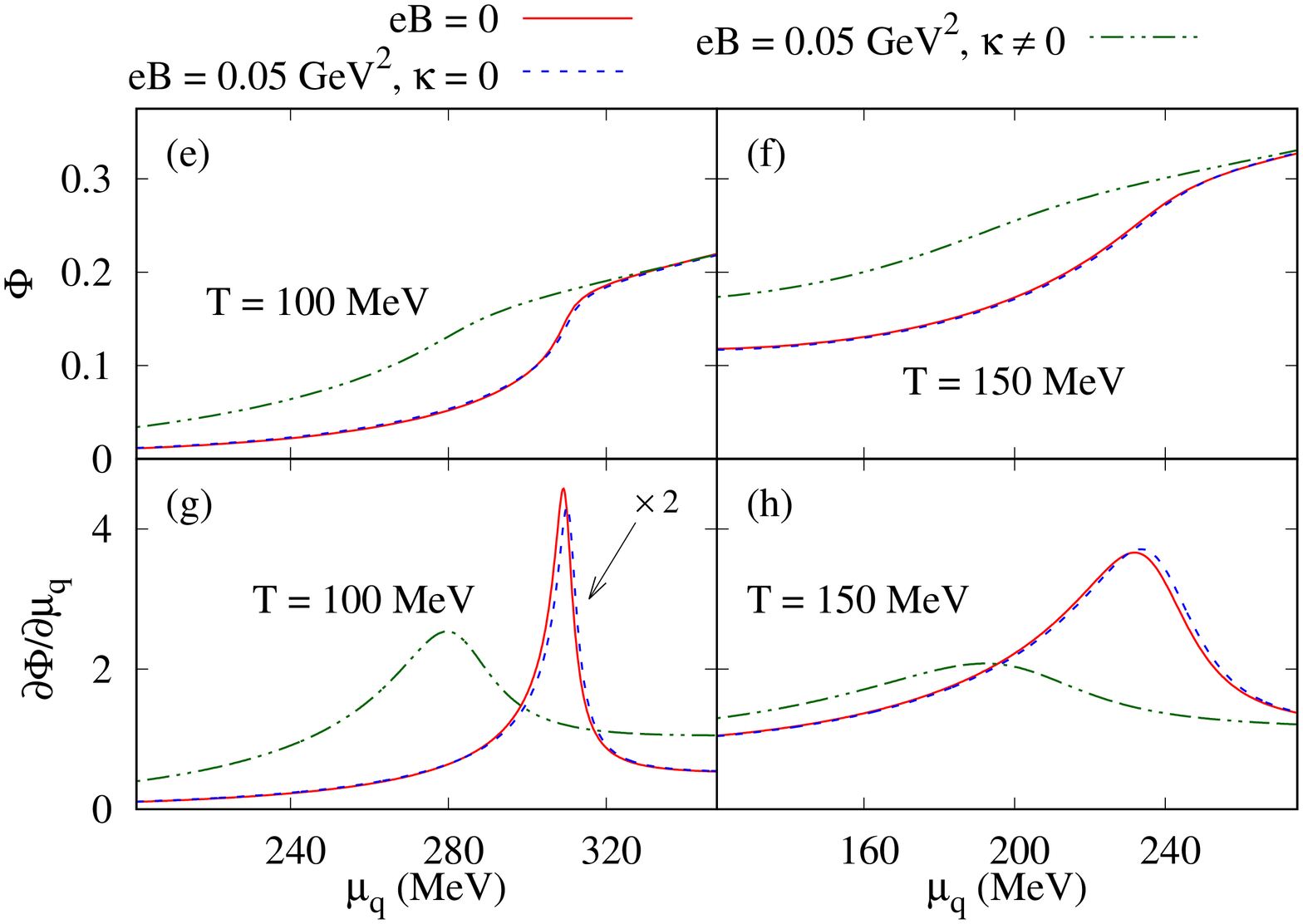}							
	\end{center}
	\caption{ Variation of $ M $, $ -\FB{\paroneder{M}{\mu_q}} $, $ \F $ and $ \FB{\paroneder{\F}{\mu_q}} $ as a function of $ \mu_q $ at different values of $ T $ and $ eB $  considering zero and non-zero values of the AMM of the quarks. In subfigures (c)  and (g) the graphs corresponding to $ \kappa = 0 $ (solid-red and dashed-blue lines) are scaled by a factor $ 1/2  $ for convenience of presentation. }
	\label{MF0eBk_vs_muq}
\end{figure}

The $ \mu_q $-dependences of  $ M $ and $ \F  $ are demonstrated in Figs.~\ref{MF_vs_Tmu} (e)-(h) for temperatures $ T=100, 125{~\rm and}~150 $ MeV respectively. In Figs.~\ref{MF_vs_Tmu} (e) and (f) the variation of $ M $ as function of $ \mu_q$  is shown and it can be seen that for a particular temperature, $ M $ remains almost constant up to certain $ \mu_q $ value and then smoothly goes to the bare quark mass limit as we increase $ \mu_q $. With increasing values of temperature $ M $ decreases for all values of $ \mu_q $ irrespective of the consideration of finite AMM of the quarks  and the transition shifts towards smaller values of $ \mu_q $.  Furthermore, as AMM of the quarks is turned on a noticeable decrease in $ M $  as $ \mu_q \rightarrow 0$ is observed from Fig.~\ref{MF_vs_Tmu} (f) for each temperature values. Also note that, in the later case, the transition from symmetry broken to restored phase occurs for lower as well as wider range of the chemical potential compared to the case when it is switched off. We will again come back to this point while discussing Figs.~\ref{MF0eBk_vs_muq} (a)-(d). In Figs.~\ref{MF_vs_Tmu} (g) and (h)  we observe that at $ \mu_q\rightarrow 0  $, the value of $ \F $ increases for higher values of temperature which follows from the fact that  as $ T $ increases the expectation value of the Polyakov loop also increases as can be seen from Fig.~\ref{MF_vs_Tmu} (c) and (d). Inclusion of AMM of the quarks hinders the rapid change in $ \F $ at higher values of $ \mu_q $ which will be more clear when we discuss the results for $ \FB{\paroneder{\F}{\mu_q}} $ later.

From Figs.~\ref{MF_vs_Tmu}(e)-(h), it is evident by comparing $ \mu_q $-dependence of $ M $ and $ \Phi $, the order parameters for chiral and deconfinement transition respectively,  that, there is a region where the expectation value of $ \Phi $ is $ \lesssim  0.4 $ and the constituent quark mass  goes to the bare quark mass limit. This is usually referred to as quarkyonic phase~\cite{MCLERRAN200986,McLerran:2007qj,Fukushima_pNJLPD,Abuki:2008nm,Carlomagno:2019yvi}. Thus  at finite chemical potential we may find a state where the	chiral symmetry has been restored while it is still in a confined phase.  Figs.~\ref{MF_vs_Tmu}(e)-(h) also depicts the fact that the formation of a quarkyonic phase is preferable at small values of temperature (for example, see the red-solid line in sub-figures (e)-(h) is for $ T = 100  $ MeV). On top of this, when the finite values of AMM of the qurks are turned on the restoration of chiral symmetry happens at smaller values of $ T (\mu_q)$ for a fixed $ \mu_q(T) $. As a consequence, the criteria of getting a quarkyonic phase is satisfied even at larger values of $ T $ as can be seen by comparing Figs.~\ref{MF_vs_Tmu}(g) and (h). For example, notice that for $ \kappa \ne 0 $, the quarkyonic phase may exist even at $ T=150  $ MeV.

In Figs.~\ref{MF0eBk_vs_T} (a) and (b) we have shown the variation of $ M $ as a function of $ T $ for $ \mu_q =0 ~{\rm and} ~ 150 ~ {\rm MeV} $ respectively for three different cases (i) $ eB = 0 $, (ii) $ eB = 0.05 ~{\rm GeV^2},~\kappa =  0 $ and  (iii) $ eB = 0.05 ~{\rm GeV^2},~ \kappa \neq  0 $. From Fig.~\ref{MF0eBk_vs_T} (a) it is evident that as we turn on the magnetic field, $ M $ increases with respect to its value at $ eB  =0$ for all values of $ T $ and as a result the transition temperature from chiral symmetry broken to the restored  phase also increases. Since we have considered finite values of bare quark mass, the pseudo-chiral transition temperature can be defined as the temperature ( $ T_C ^\chi$) for which $ M $ has the highest change.
Now from Fig~\ref{MF0eBk_vs_T} (c), one can observe that the peak of $ -\FB{\paroneder{M}{T}} $ has shifted marginally towards the higher values of temperature when finite value of $ eB $ is considered (the blue-dashed line), which is evident from the inset plots. This indicates \textit{magnetic catalysis} (MC), which implies that the finite values of the  magnetic field results in the enhancement of chiral condensates $ \pbarpasi $. On the contrary, an opposite behaviour is observed when we include non-zero values of AMM of the quarks in presence of the background magnetic field and the transition temperature ($ T_C ^\chi$) decreases. This feature is also evident from Fig~\ref{MF0eBk_vs_T} (c) where the peak of $ -\FB{\paroneder{M}{T}} $ shifts towards the  lower values of $ T $ (dash-dot green curve), confirming \textit{inverse magnetic catalysis} (IMC). At finite values of the quark chemical potential we observe further decrease in $ T_C ^\chi$ values which is evident from Figs.~\ref{MF_vs_Tmu} (a) and (b) but the overall nature remains same. However, it is interesting to note that as we increase $ \mu_q $, the magnitude of $ -\FB{\paroneder{M}{T}} $ becomes larger and the peak becomes narrower. Thus, one can conclude that as we increase $ \mu_q $ the rate of change of $ M $ increases and the transition occurs at smaller range of temperatures. 

In Figs.~\ref{MF0eBk_vs_T} (e) and (f) we present the variation of $ \F $ as  a function of $ T $ for $ \mu_q =0 ~{\rm and} ~ 150 ~ {\rm MeV} $ respectively for three different cases (i) $ eB = 0 $, (ii) $ eB = 0.05 ~{\rm GeV^2},~\kappa =  0 $ and  (iii) $ eB = 0.05 ~{\rm GeV^2},~ \kappa \neq  0 $. From both the plots (Figs.~\ref{MF0eBk_vs_T} (e) and (f)) it can be seen that when only the effect of background magnetic field is taken into consideration the change in $ \F $ as a function of temperature is practically negligible. However with the inclusion of AMM of the quarks the transition temperature decreases substantially. This fact is also seen from Figs.~\ref{MF0eBk_vs_T} (g) and (h) where one can observe the shift of $ \FB{\paroneder{\F}{T}}$ peaks towards the lower values of $ T  $ when we consider non-zero AMM of the quarks (dash-dot-green line) in both the figures.  Finite values of quark chemical potential results in the following noticeable effects. Firstly, the magnitude of $ \F $ decreases as compared to $ \mu_q = 0 $ case and the difference becomes larger with increasing values of temperature. Secondly, as the magnitude of $ \F $ is lower, the rate of change of $ \F $ with the variation of temperature (i.e. $ \FB{\paroneder{\F}{T}} $) is also small in magnitude (results in broadening). Finally, the peak of $ \FB{\paroneder{\F}{T}} $ is slightly left shifted as compared to the $ \mu_q =0$ scenario.

From Figs.~\ref{MF0eBk_vs_T}(c)-(f), it is evident that the critical temperatures for chiral and deconfinement transition do not coincide. 
		This is expected in local PNJL approach~\cite{Ratti} which has been considered in this work, irrespective of the form of Polyakov potential~\cite{Ratti3,Abuki:2008nm}. 
		But there are many important modifications of this model available in the literature e.g. inclusion of the effect of the $ SU(3) $ measure with a 
		Vandermonde term such that the Polyakov loop always remains in the domain $ [0, 1] $~\cite{Ghosh:2007wy}. Lattice QCD simulation~\cite{Aoki2006,Fukugita} has 
		confirmed that these two transitions occur almost at the same temperature. It was proposed in~\cite{TSasaki_epNJL} that this coincidence can be ensured through 
		a strong correlation or entanglement between the chiral condensate $ (\sigma) $ and the  expectation value of $ (\Phi) $  within the PNJL model, which is referred 
		to as entanglement PNJL (EPNJL). Moreover, using non-local four fermion interaction~\cite{Ripka:1997zb}, one can extend NJL model further with the intention to 
		provide a more realistic effective approach to QCD (see ~\cite{GomezDumm:2017jij,Noguera:2008cm,Contrera:2009hk} and references therein for details). 

In Figs.~\ref{MF0eBk_vs_muq} (a) and (b) the variation of $ M $ as a function of $ \mu_q $ for $ T =100 ~{\rm and} ~ 150 ~ {\rm MeV} $ respectively is depicted for three different cases (i) $ eB = 0 $ (ii) $ eB = 0.05 ~{\rm GeV^2},~\kappa =  0 $ and  (iii) $ eB = 0.05 ~{\rm GeV^2},~ \kappa \neq  0 $. From Fig.~\ref{MF0eBk_vs_muq} (a) it can be seen that the presence of non-zero background magnetic field increases the values of $ M $ for the whole range of $ \mu_q $ and consequently the transition temperature from chiral symmetry broken to restored phase also increases, which is evident from Fig.~\ref{MF0eBk_vs_muq} (c) where the peak of $ -\FB{\paroneder{M}{\mu_q}} $ is shifted towards the higher values of $ T $ indicating MC. On the other hand, inclusion of finite AMM of the quarks leads to a substantial decrease in $ M$ for all the  values of $ \mu_q $ and as a results the transition temperature decreases which is evident from Fig.~\ref{MF0eBk_vs_muq} (c). This phenomena can be classified as IMC. However, notice that the peaks for cases (i) and (ii) are much higher and sharper compared to case (iii) (we have scaled down $ -\FB{\paroneder{M}{\mu_q} }$ in Fig.~\ref{MF0eBk_vs_muq} (c) by a factor of $ 2 $).   Furthermore, as we increase the temperature from $ 100 $ to $ 150 $ MeV, we observe a broadening of the peaks. This is expected from the discussions of Figs.~\ref{MF_vs_Tmu} (e) and (f), where we have already pointed out that  the transition from symmetry broken to restored phase occurs for lower as well as over a wider range of chemical potential compared to the case when AMM of the quarks is switched off.

In Figs.~\ref{MF0eBk_vs_muq} (e) and (f)  the variation of $ \F $ as function of $ \mu_q $ is displayed for $ T =100 ~{\rm and} ~ 150 ~ {\rm MeV} $ respectively for three different cases (i) $ eB = 0 $, (ii) $ eB = 0.05 ~{\rm GeV^2},~\kappa =  0 $ and  (iii) $ eB = 0.05 ~{\rm GeV^2},~ \kappa \neq  0 $. Here we observe that the presence of the background magnetic field affects the $ \mu_q $-dependence of $ \F $ marginally. However, as we include AMM of the quarks a noticeable difference can be seen. In both the cases this leads to a decrease in the deconfinement transition temperature as compared to the zero AMM case. The results shown in Figs.~\ref{MF0eBk_vs_muq} (g) and (h) further confirm our observations. Note that, the behaviour of $ \FB{\paroneder{\F}{\mu_q}}  $  are quite similar to that discussed in the last paragraph. 

\begin{figure}[h]
	\begin{center}
		\includegraphics[angle=-90, scale=0.7]{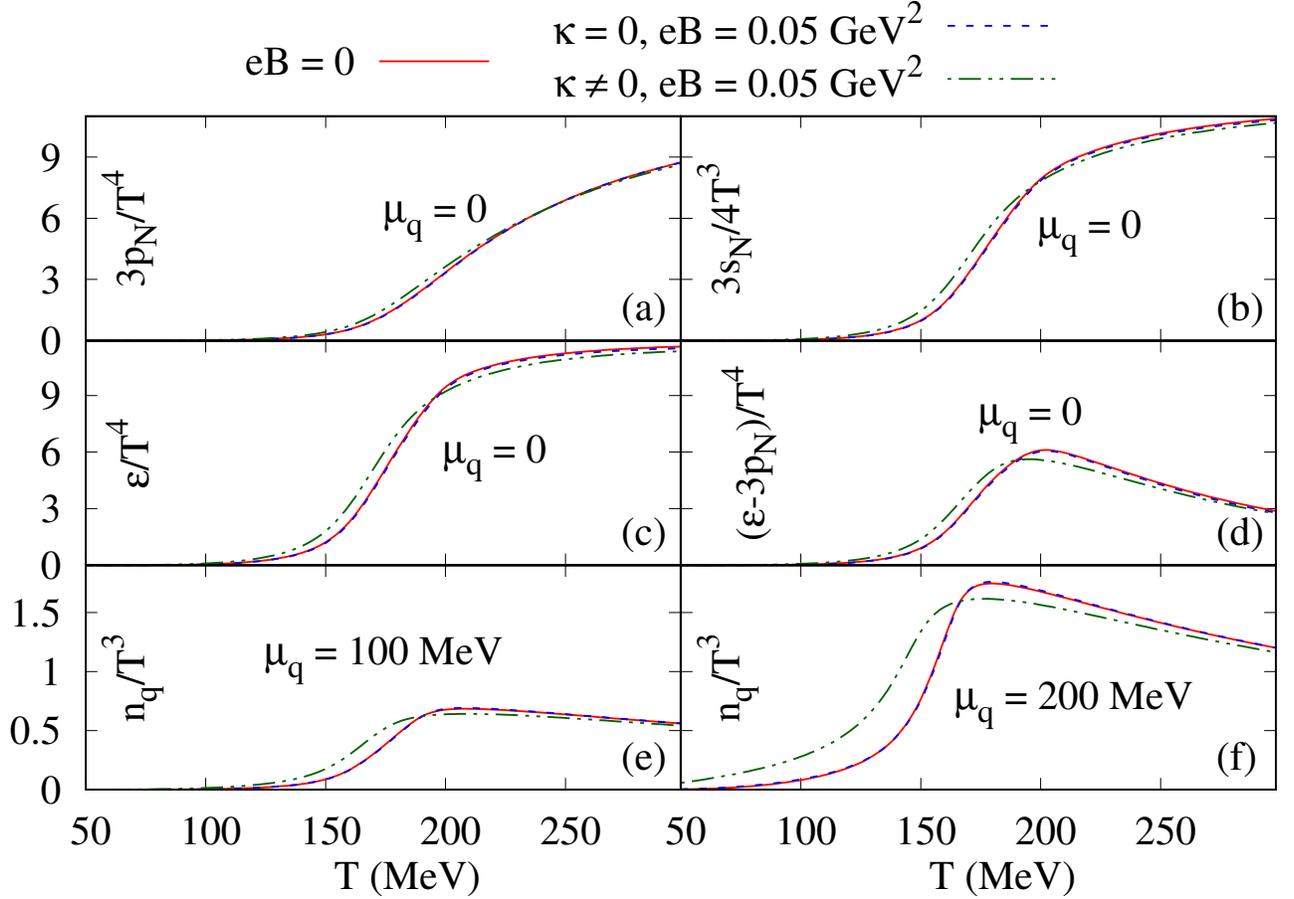}				
	\end{center}
	\caption{ Variation of scaled $ \varepsilon, p, s $, $ (\varepsilon - 3p) $ and $ n_q $ as function of $ T $  for different values of $eB $ and $ \kappa $.}
	\label{eps_vs_T}
\end{figure}

In Figs.~\ref{eps_vs_T} (a), (b) and (c) we plot the scaled pressure, entropy and energy density  respectively as a function of temperature at zero chemical potential. The scaling  is done in the usual fashion:
\begin{equation}
X_N\FB{T,\mu_q,eB} \equiv X\FB{T,\mu_q,eB} - X \FB{0, \mu_q, eB}
\end{equation} 
where $ X\in \SB{p,s,\varepsilon } $ and is divided by different powers of $ T $ to make the quantities dimensionless. Since the transition from the symmetry broken to the restored phase, as previously discussed,  is a rapid crossover,  the pressure, entropy and the energy densities are continuous functions of the temperature. The overall behaviour is similar in the three curves: a sharp increase in the vicinity of the transition temperature followed by a tendency to saturate. Finite values of  magnetic field i.e. $ eB=0.05 ~{\rm GeV^2} $ hardly brings any noticeable change in the above mentioned quantities. However, when we include the AMM of the quarks, the transition temperature shifts towards the  lower values of temperature which is expected from the previous discussions. From Fig.~\ref{eps_vs_T} (d) it is evident that the non-zero values of AMM of the quarks shift the peak of the interaction measure $ (\varepsilon - 3p)_N $ towards lower temperature values.

The reduced quark number density $ n_q/T^3 $ is presented in  Figs.~\ref{eps_vs_T} (e) and (f) as a function of temperature for $ \mu_q = 100 $ and $ 200 $ MeV respectively. The behaviour of $ n_q $ can be explained following~\cite{Ratti,SasakipNJL}. Let us first concentrate on Fig.~\ref{eps_vs_T} (e) where we have considered the  following three cases: (i) $ eB = 0 $, (ii) $ eB = 0.05 ~{\rm GeV^2},~\kappa =  0 $ and  (iii) $ eB = 0.05 ~{\rm GeV^2},~ \kappa \neq  0 $ at $ \mu_q  = 100 MeV $. In each case, for temperatures below the transition, the interaction with the effective gluon field leads to suppressions of one and two-quark contributions to the density. As a result, the three-quark states become more dominant. Thus, we observe a strong suppression of the quark density below transition. However, for temperatures above the transition, this suppression is  less effective. But as $ \F $ is still less than  unity (see Fig.~\ref{MF_vs_Tmu} (c) and (d)) a marginal suppression can be observed compared to the quark density of a free gas (which is also observed in NJL model). In case (ii) we observe a similar qualitative behaviour of $ n_q/T^3 $. This is because when we turn on the background magnetic field, at high values of temperature the difference in $ M $  is almost negligible (see Fig.~\ref{MF0eBk_vs_T} (a)) compared to the zero field case. Furthermore, at low values of temperature the one and two-quark contributions remain strongly suppressed as finite $ eB $ strengthen the chiral condensate (as a result $ M $ increases).  However, when we include AMM of the quarks, $ M $ has sufficiently low magnitude even at high values of $ T $ compared to cases (i) and (ii). Thus the suppression of $ n_q $  is larger at high temperature in case (iii). On the other hand, from Figs.~\ref{MF0eBk_vs_T} (a) and (e) one can observe that the magnitude of $ M $ ($ \F $) is smaller (higher) at lower values of temperature when AMM of the quarks is taken into consideration. As a result, the one and two-quark states become dominant at lower values of temperature in contrast to the other two cases and thus the transition occurs at lower values of $ T $. Now, in Fig.~\ref{eps_vs_T} (f) we have used higher values of $ \mu_q $ which leads to the decrease in transition temperature  at much faster rate as discussed earlier (see Figs.~\ref{MF0eBk_vs_T} (a)-(d)). This explains the rise of $ n_q $ at lower values of $ T $ compared to the previous one.

\begin{figure}[h]
	\begin{center}
		\includegraphics[angle=-90, scale=0.3]{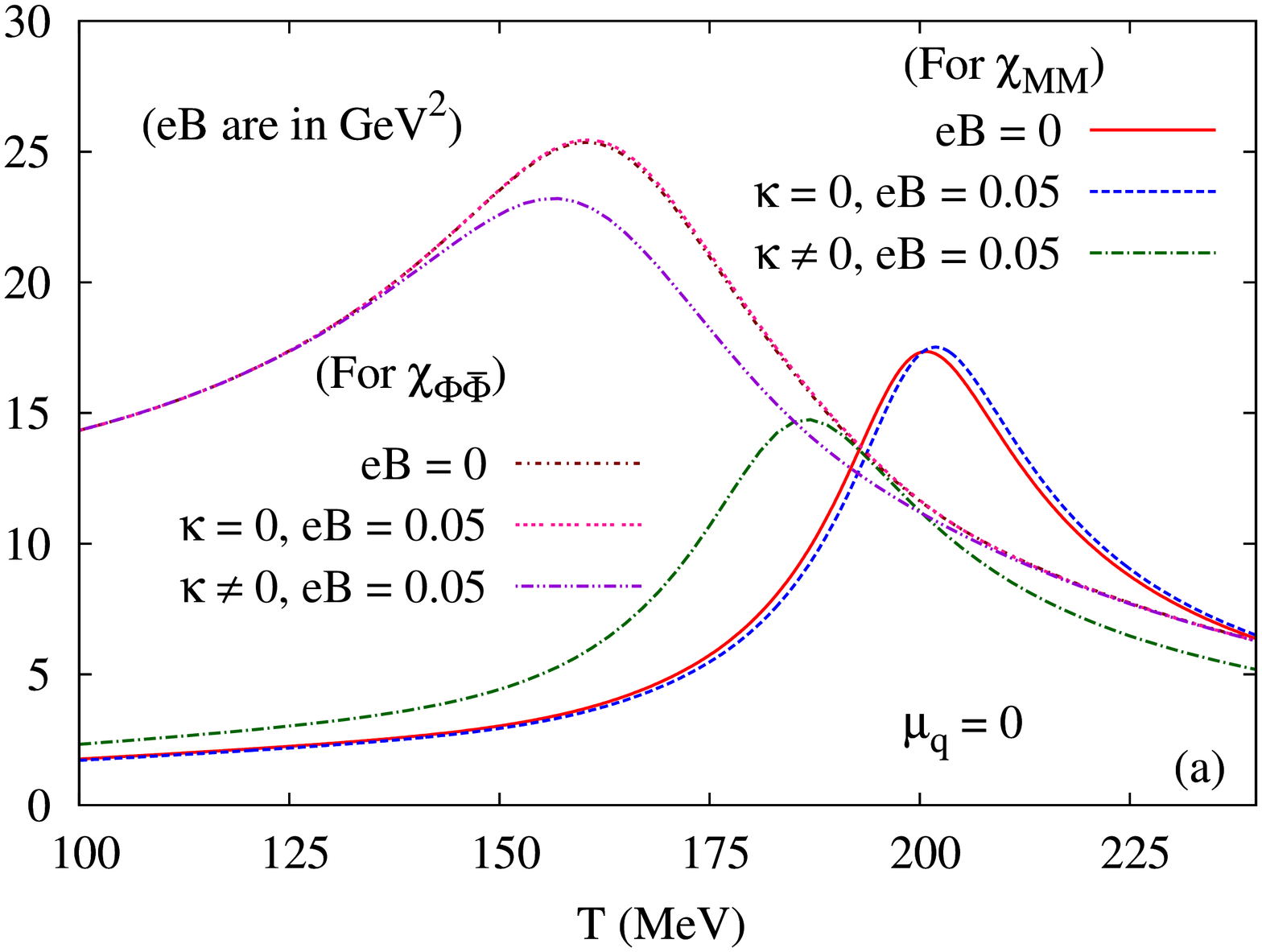}
		\includegraphics[angle=-90, scale=0.3]{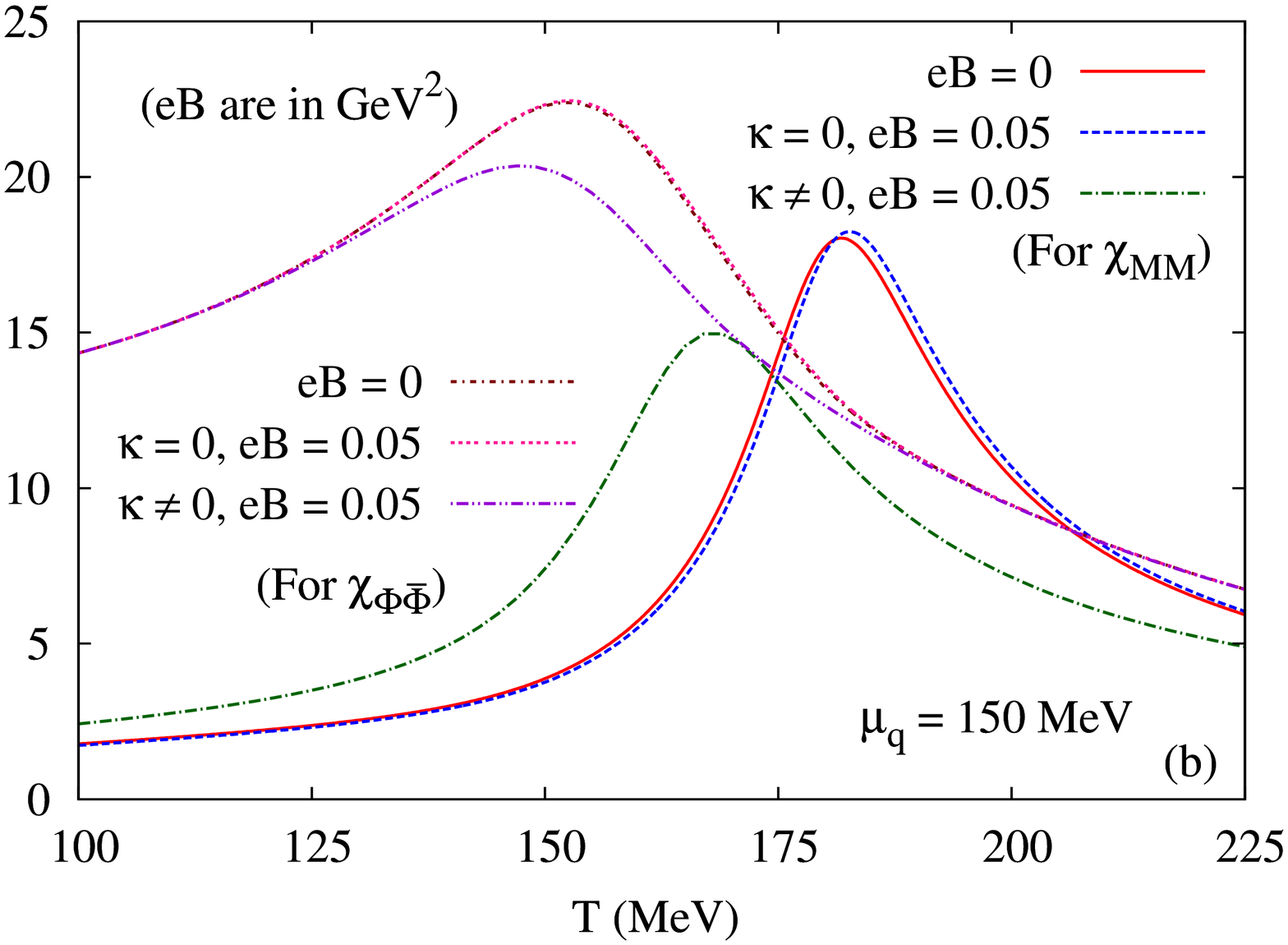}					
	\end{center}
	\caption{ Variation of $ \chi_{MM} $ and $ \chi_{\F\Fb } $ as function of $ T $  for different values of $ \mu_q $, $eB $ and $ \kappa  $. }
	\label{chiMM_vs_T_eBk}
\end{figure}

We now focus on the results for different susceptibilities. As discussed earlier, susceptibilities associated with $ M $ and $ \F  $ are the effective fields, which show signals of phase transitions and can be considered as order parameters  for chiral and deconfinement transitions respectively. Now the off-diagonal susceptibility $ \chi_{\F\Fb} $ is $ Z(3) $ invariant but the diagonals are not. This property makes $ \chi_{\F\Fb} $ a good candidate to study the deconfinement transitions in PNJL model~\cite{SasakipNJL}. Furthermore, results for quark number susceptibility ($ \chi_q $), specific heat ($ C_V $) and velocity of sound ($ c_s $) are also obtained using Eqs.~\eqref{qnumsus}~\eqref{cv} and \eqref{velsound} respectively. All these results are shown for the following three cases: (i) $ eB = 0 $, (ii) $ eB = 0.05 ~{\rm GeV^2},~\kappa =  0 $ and  (iii) $ eB = 0.05 ~{\rm GeV^2},~ \kappa \neq  0 $. Figs~\ref{chiMM_vs_T_eBk} (a) and (b) show the $ T $-dependence of $ \chi_{MM}  $ and $ \chi_{\F\Fb } $ at $ \mu_q = 0 $ and $ 150  $ MeV respectively for the three cases previously mentioned. It is evident that when only the presence of background magnetic field is taken into consideration $ T_C^\chi  $ moves towards the higher values of temperature implying MC. On the contrary, inclusion of AMM of the quarks results in decrease in $ T_C^\chi  $ which can be identified as IMC. Clearly, inclusion of AMM of the quarks decreases the deconfinement  transition temperature ($ T_C^d $) substantially which is evident from both the plots. Now, for finite values of $ \mu_q $ we notice that there is an overall decrease in $ T_C^\chi $ and $ T_C^d $ but the qualitative nature remains similar. These results are in agreement with our observations while discussing Fig.~\ref{MF0eBk_vs_T}. Note that as we increase the quark chemical potential, the peak position of the chiral and Polyakov loop susceptibilities approach each other, as seen in~\cite{SasakipNJL}. The perfect coincidence of the chiral and deconfinement transitions are lost due to our choice of $ T_0 = 190  $ MeV, following the argument presented in~\cite{Ratti}. The similar behaviour is also reported in~\cite{Costa}.

\begin{figure}[h]
	\begin{center}
		\includegraphics[angle=-90, scale=0.3]{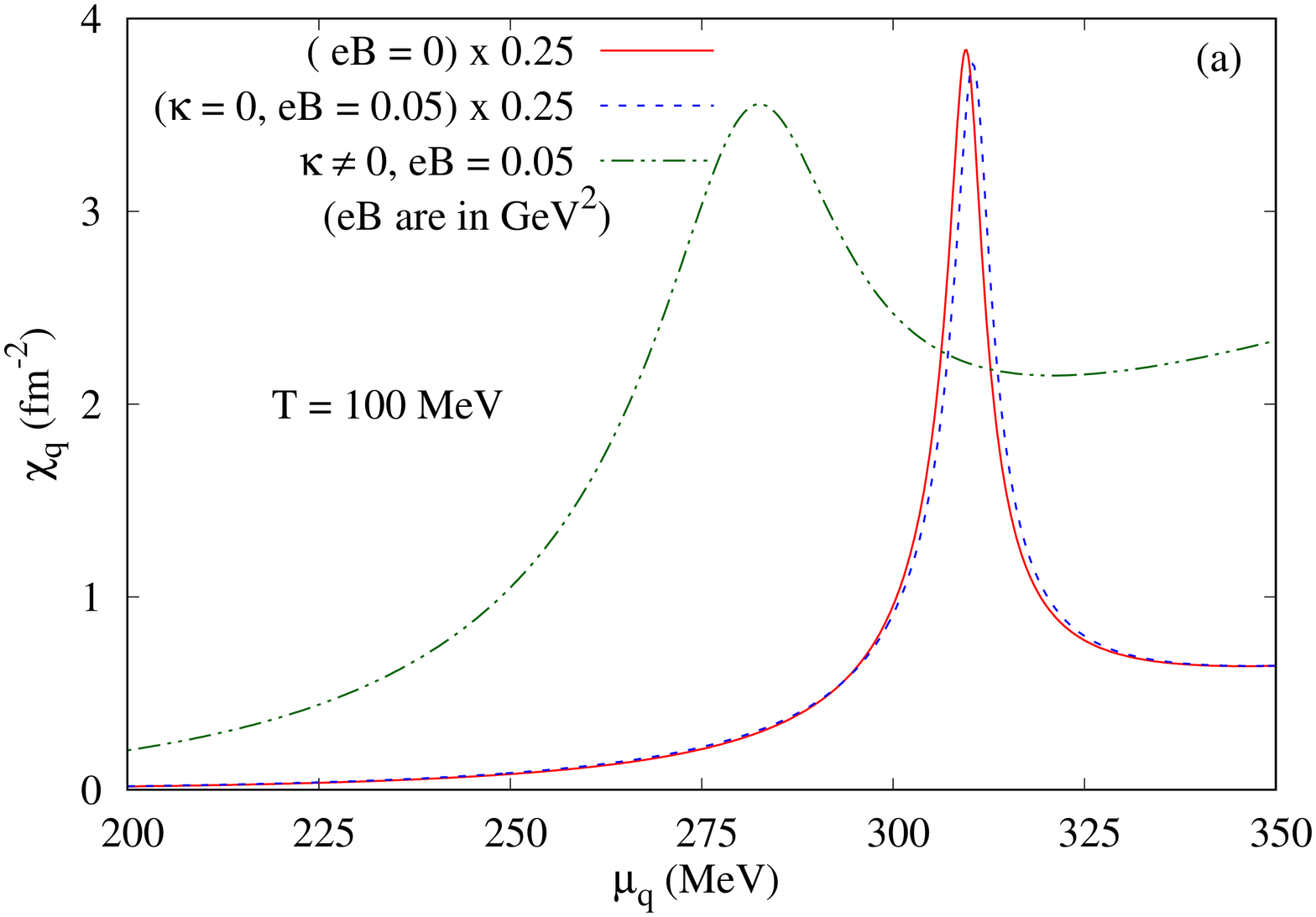}
		\includegraphics[angle=-90, scale=0.3]{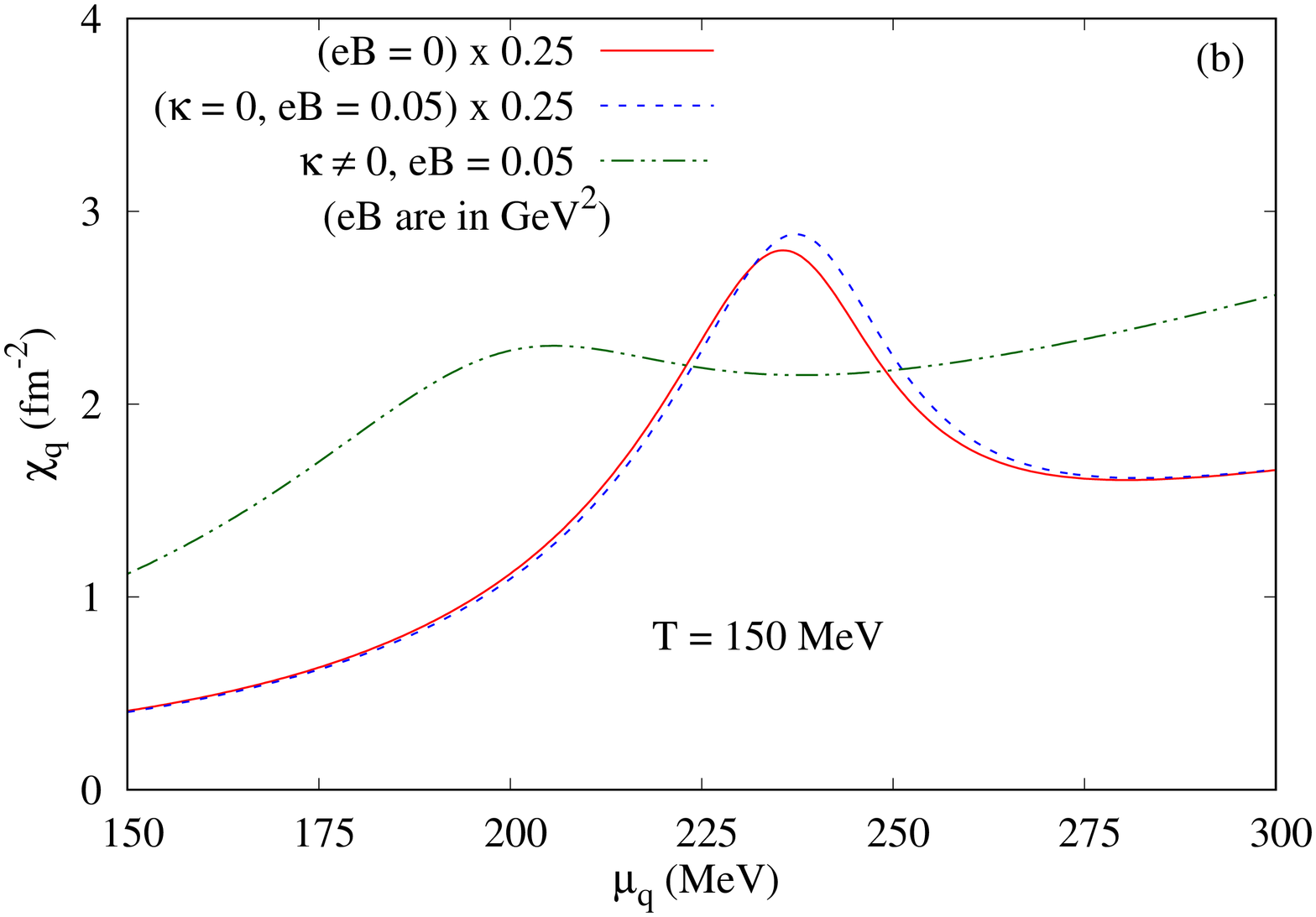}					
	\end{center}
	\caption{ Variation of $ \chi_q $ as function of $ \mu_q $  for different values of $ T $, $eB $ and $ \kappa  $.  The graphs corresponding to $ \kappa = 0 $ (solid-red and dashed-blue lines) are scaled by a factor $ 1/4 $ for convenience of presentation. }
	\label{Xq_vs_mu}
\end{figure}

In Figs.~\ref{Xq_vs_mu} (a) and (b) we have shown $ \chi_q  $ as a function of $ \mu_q $  at two different temperatures. As expected, we get IMC (MC) when we consider finite values of AMM of the quarks in presence of the background magnetic field (AMM of the quarks are switched off). One can make direct correspondence between these two plots with the results shown in Figs.~\ref{MF0eBk_vs_muq} (c) and (d). Absence of any discontinuity in the curves implies that  the transition is  crossover.  

\begin{figure}[h]
	\begin{center}
		\includegraphics[angle=-90, scale=0.3]{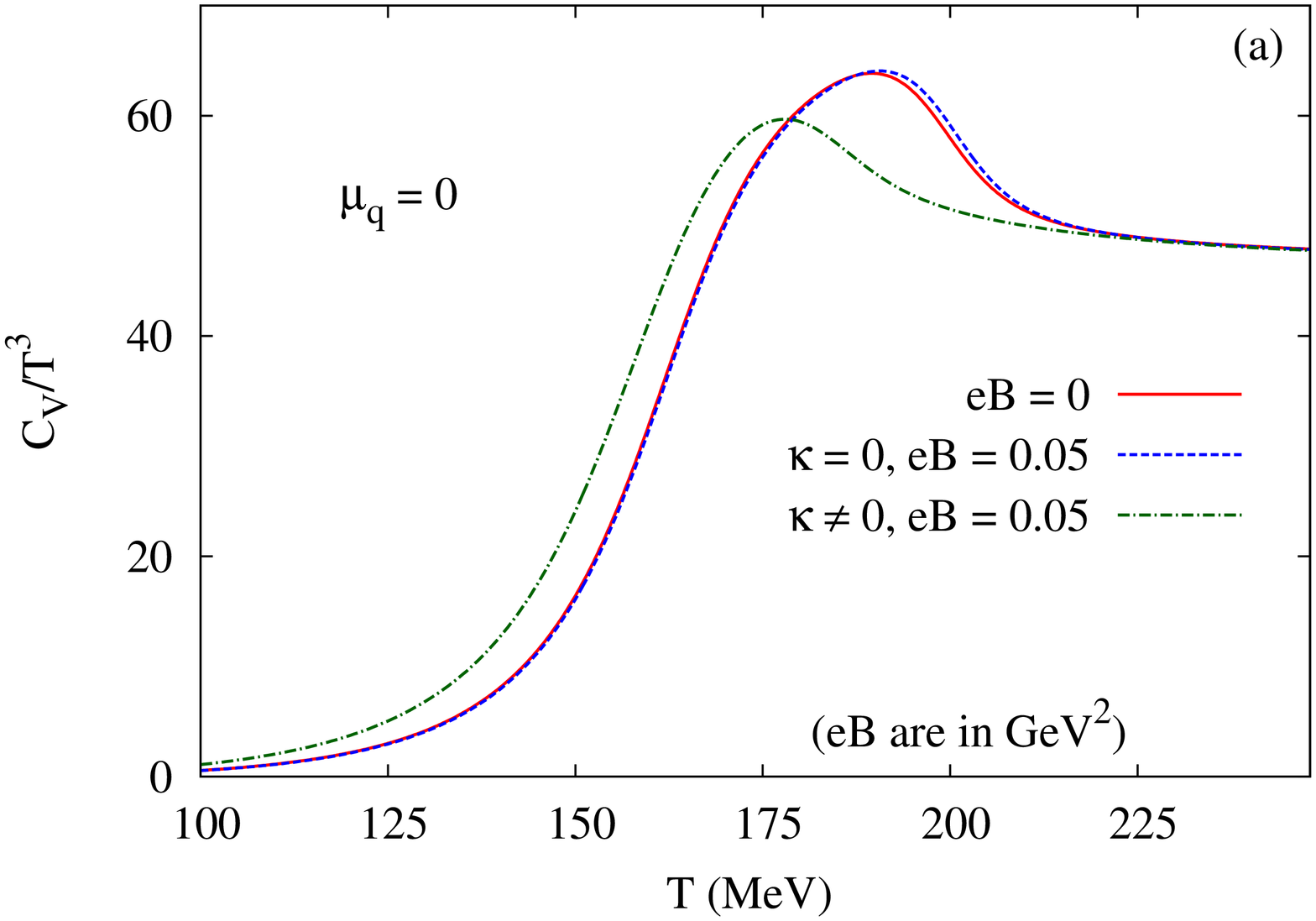}
		\includegraphics[angle=-90, scale=0.3]{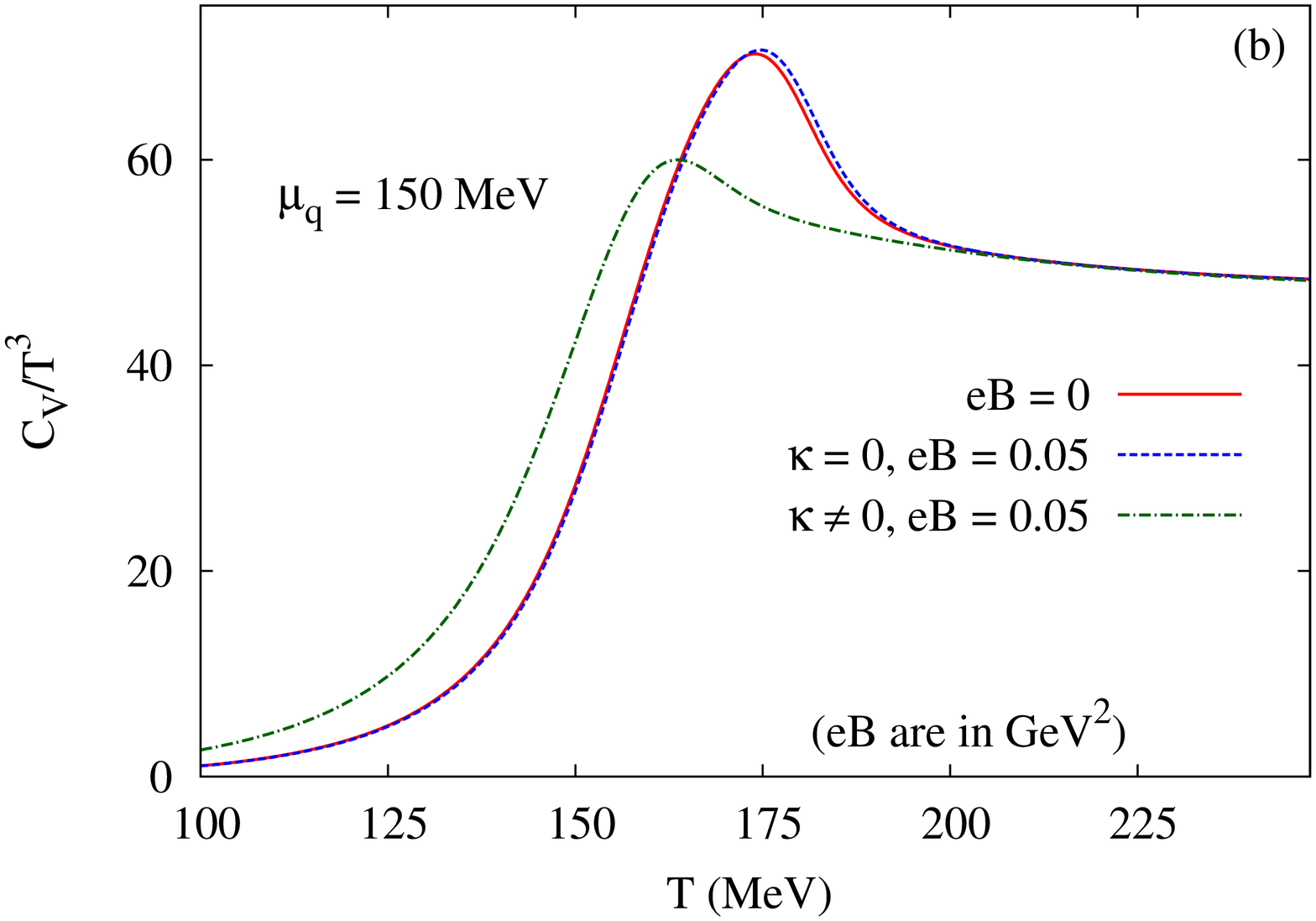}					
	\end{center}
	\caption{ Variation of $ C_V/T^3 $ as function of $ T $  for different values of $ \mu_q $, $eB $ and $ \kappa  $. }
	\label{CV_vs_T}
\end{figure}

 In Figs.~\ref{CV_vs_T} (a) and (b) we have plotted $ C_V/T^3 $ as a function of $ T $ for zero and finite values of $ \mu_q $. It is observed that, in both occasions, $ C_V $ grows with increasing temperature and reaches a peak at the transition point and decreases sharply for a short range of temperature. Thereafter it slowly saturates to a value slightly lower than the ideal gas value at high temperature. Inclusion of AMM of the quarks at non-zero background magnetic field consequently shifts the peak towards lower values of $ T $. At finite $ \mu_q $, there are overall leftward shifts of all the plots, but the qualitative natures remain the same.

\begin{figure}[h]
	\begin{center}
		\includegraphics[angle=-90, scale=0.225]{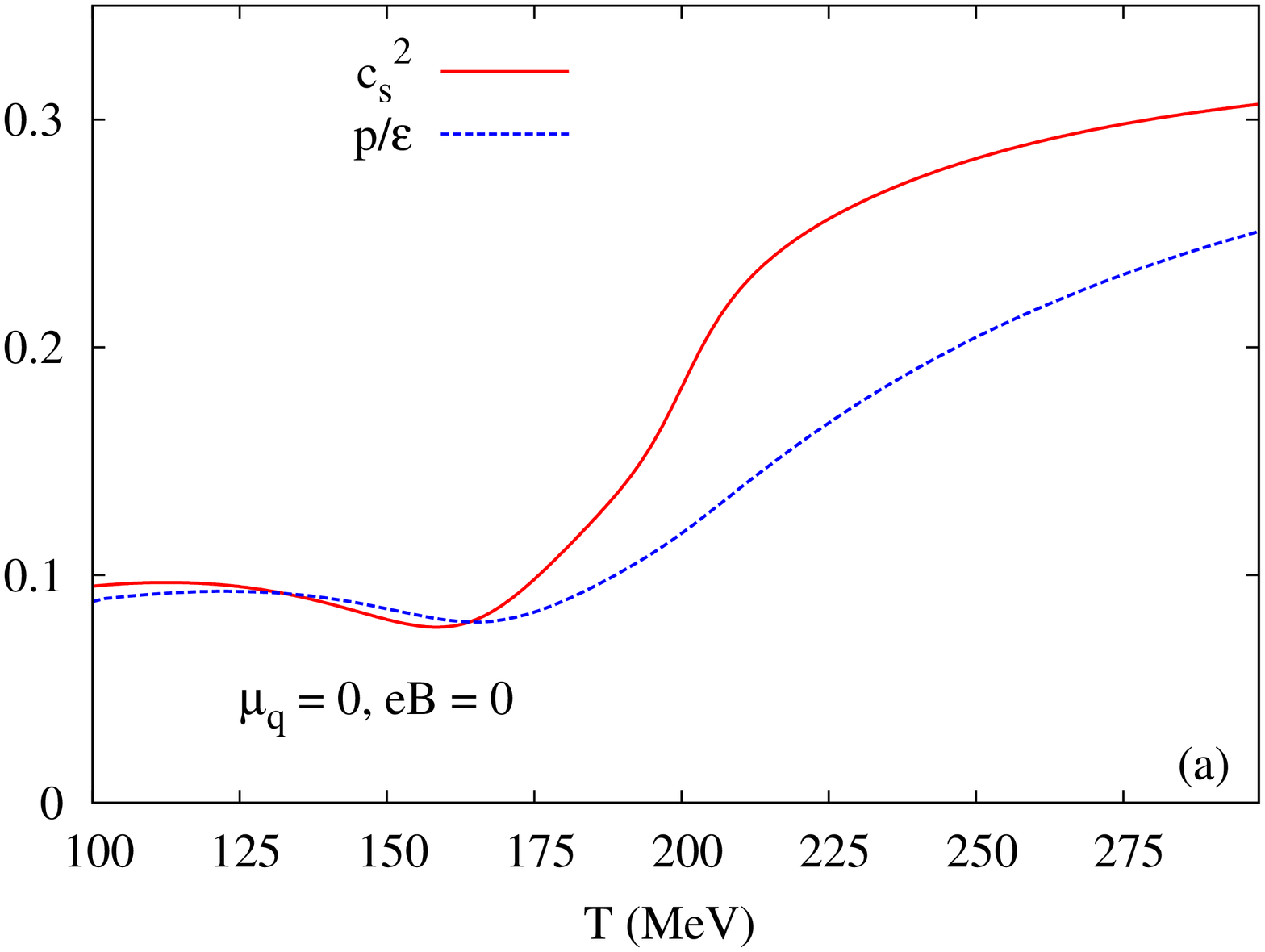}
		\includegraphics[angle=-90, scale=0.225]{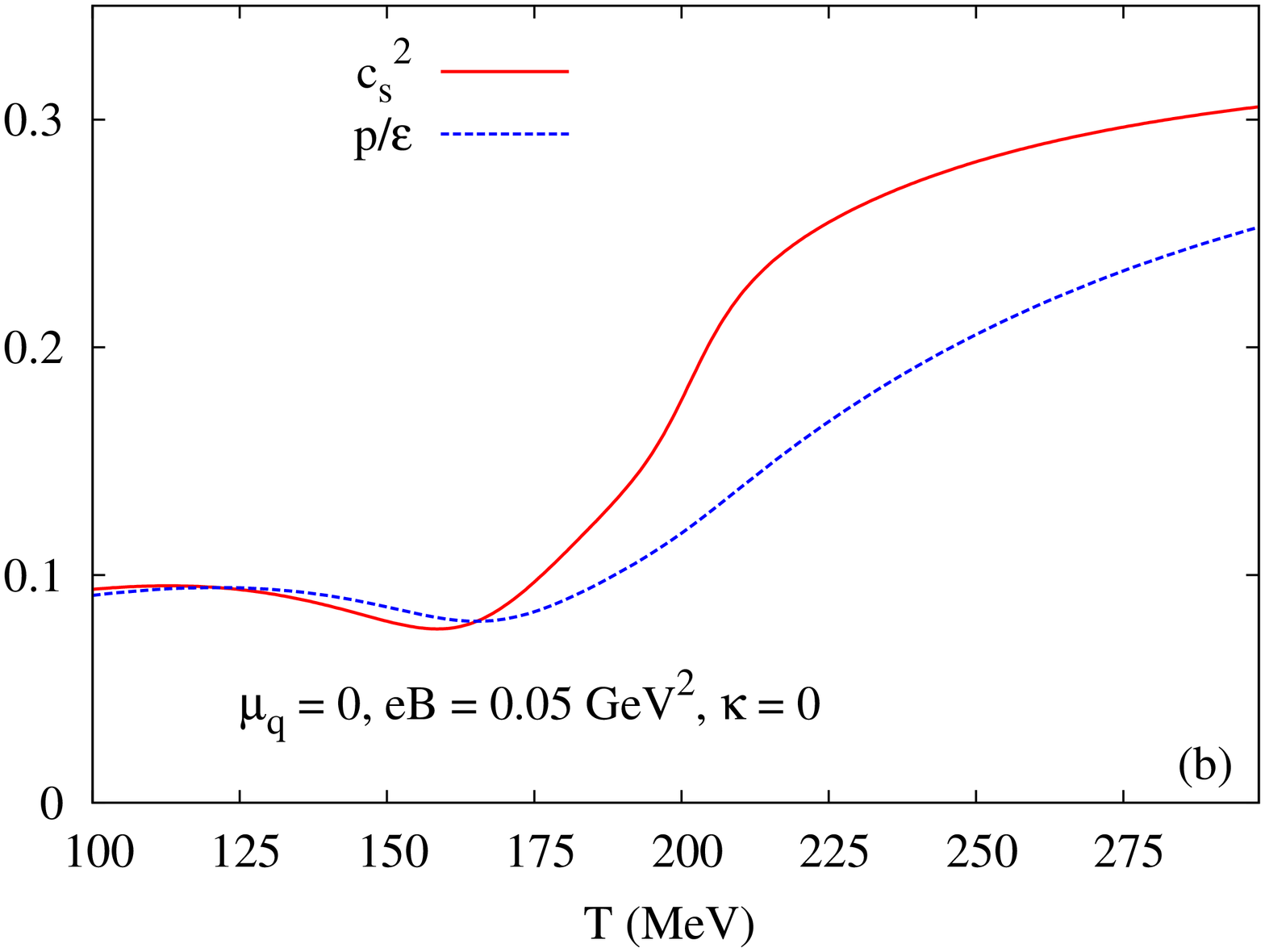}					
		\includegraphics[angle=-90, scale=0.225]{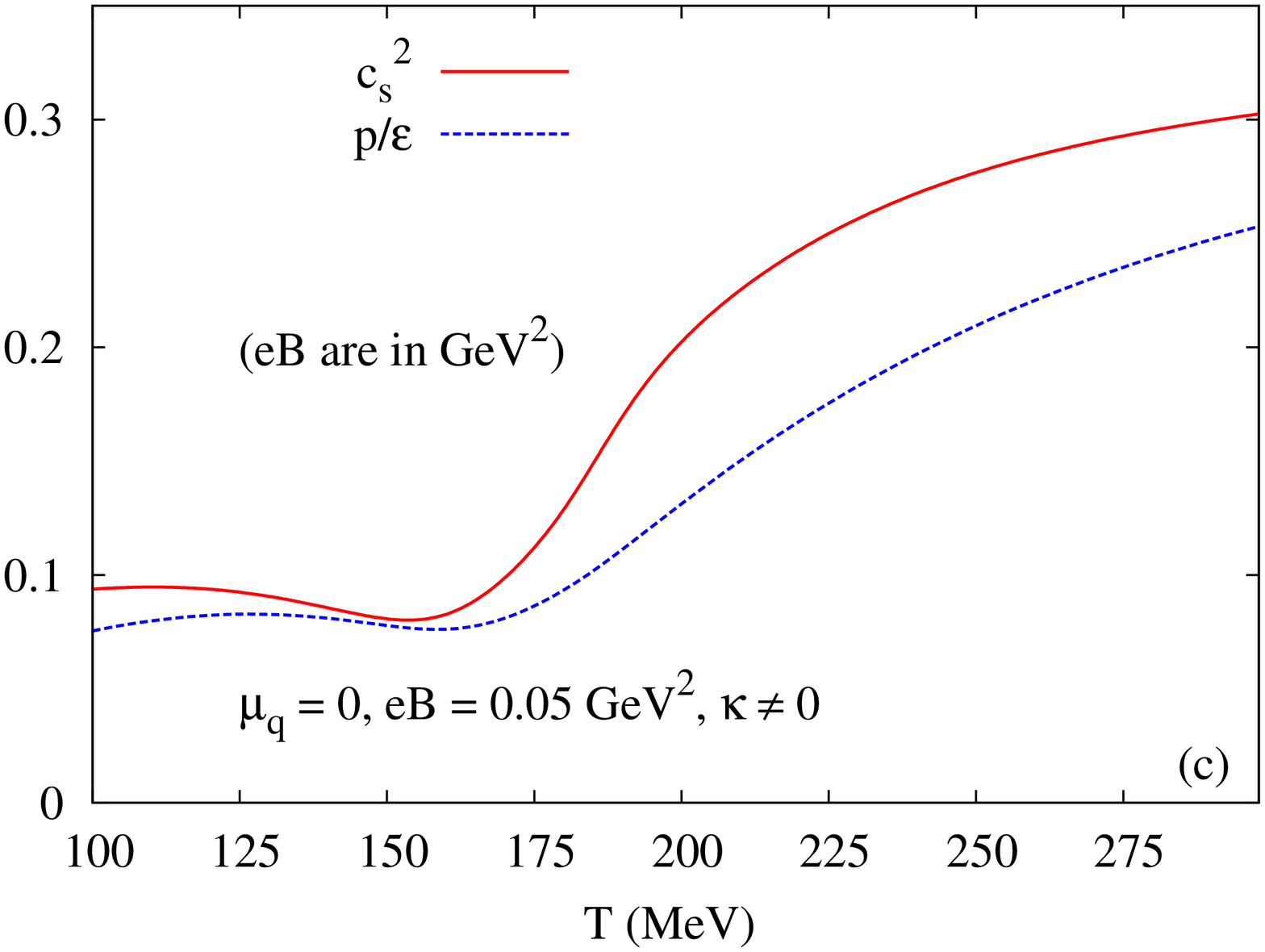}					
	\end{center}
	\caption{ Variation of $ c_s^2 $ and $ p/\varepsilon  $ as function of $ T $  for different values of $eB $ and $ \kappa $ at $ \mu_q  = 0 $. }
	\label{CSp_vs_T}
\end{figure}

In Figs.~\ref{CSp_vs_T} we have shown the variation of $ c_s^2 $ and $ p/\varepsilon  $ as a function of temperature at $ \mu_q  =0$ for three different cases. As defined in Eq.~\eqref{velsound}, denominator of $ c_s^2 $ is nothing but $ C_V $, a minima is expected near the transition. In all the plots, one such pronounced dip can be seen. After the crossover, release of the new degrees of freedom results in rapid increase of the speed of sound, which is evident from all the plots. The minimum of the speed of sound, known as the softest point, may be an important indicator of the transition observed in heavy-ion collisions ~\cite{Hung}. As a consequence of incorporation of finite values of AMM of the quarks  this minima shifts towards lower values of $ T $. It is important to note that, the value of $ p/\varepsilon  $ nearly matches with $ c_s^2 $ below transitions and becomes close again as we increase the temperature. But in between, $ c_s^2 $ is distinctly greater than $ p/\varepsilon  $. (see~\cite{Ratti3,Sanjay,Abhijit} for discussions).

\begin{figure}[h]
	\begin{center}
		\includegraphics[angle=-90, scale=0.3]{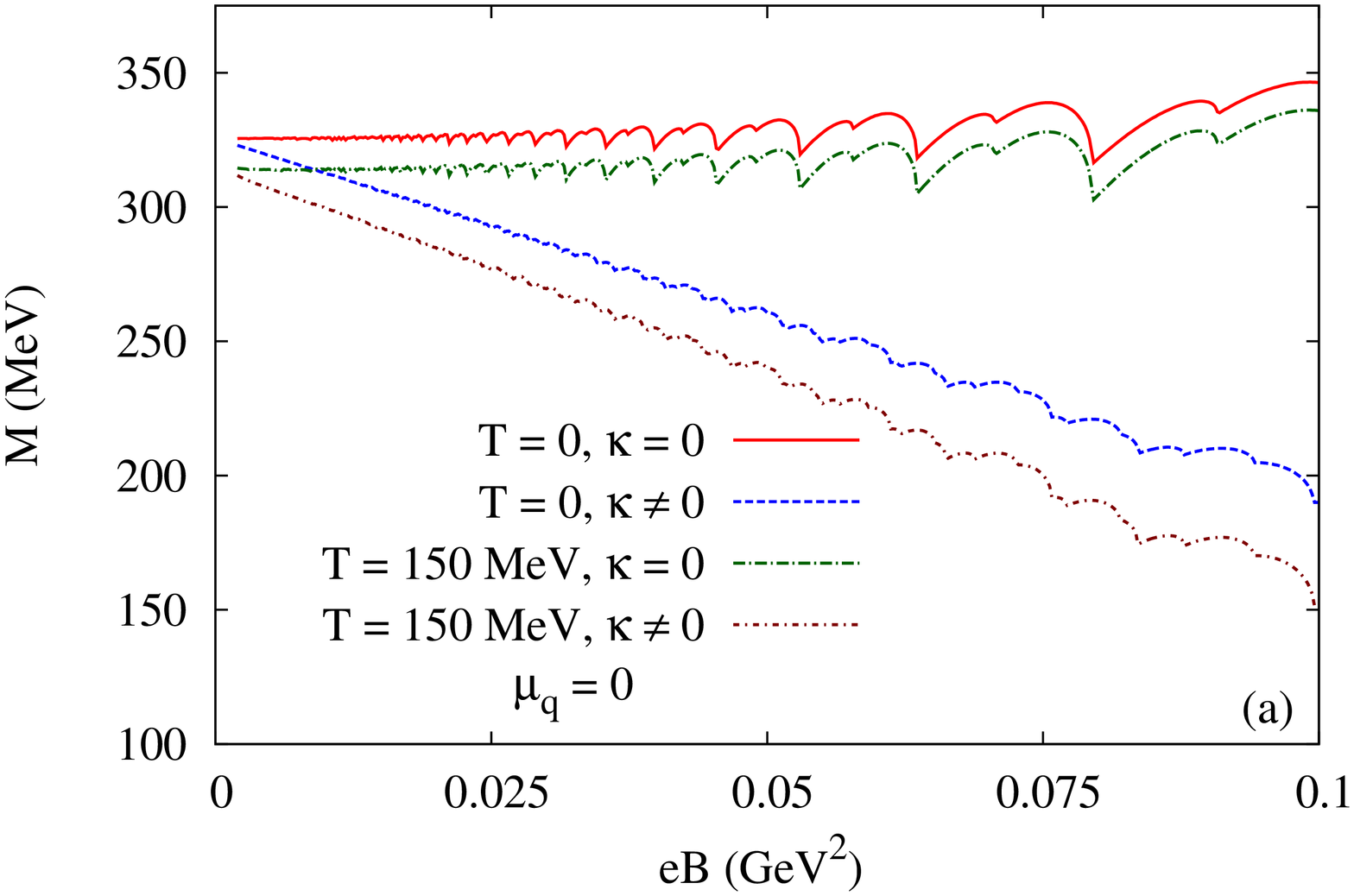}
		\includegraphics[angle=-90, scale=0.3]{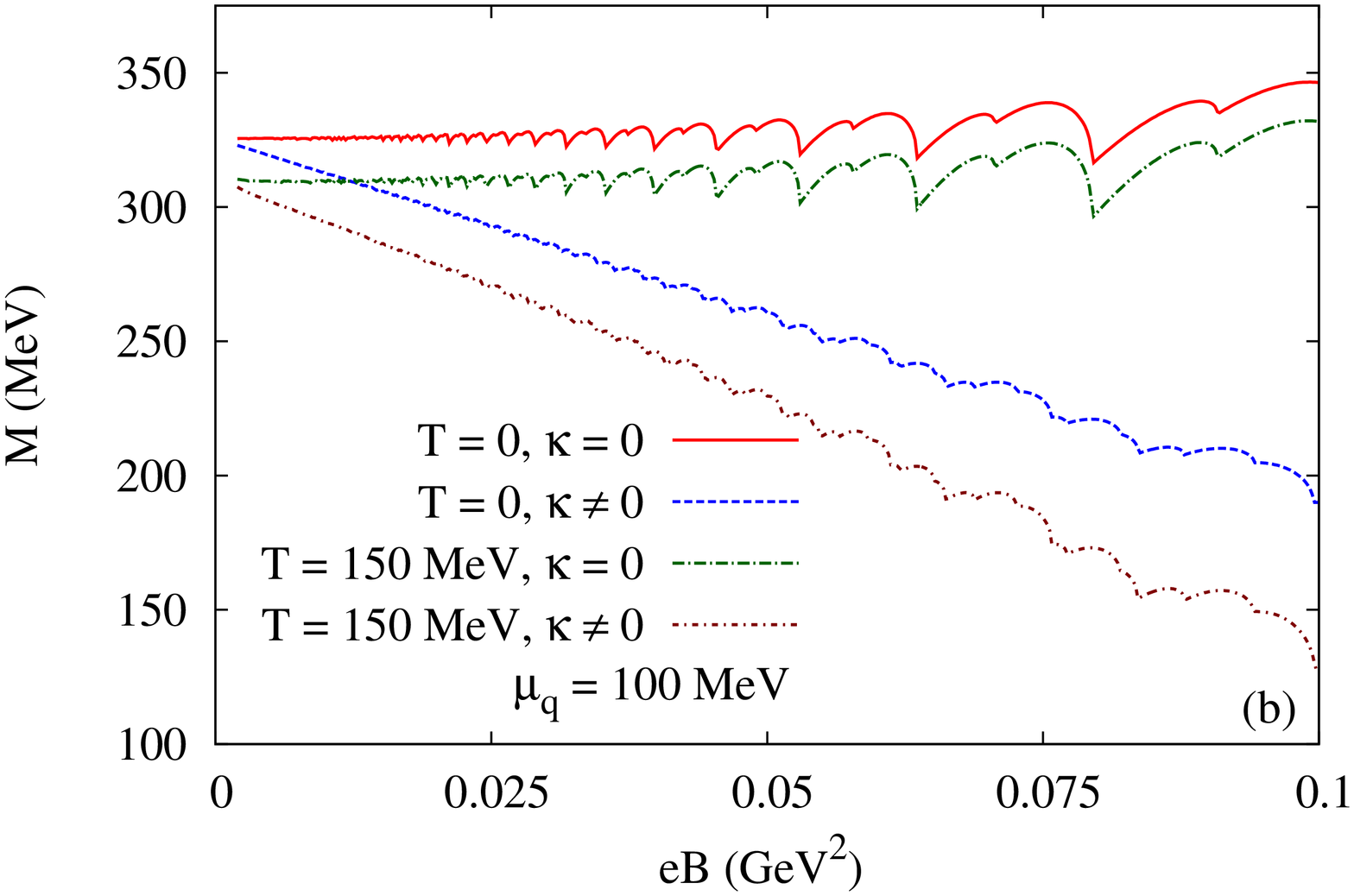}					
	\end{center}
	\caption{ Variation of $ M $ as function of $ eB $  for different values of $T $, $ \mu_q $ and $ \kappa $. }
	\label{M_vs_eB}
\end{figure}

In Figs.~\ref{M_vs_eB}(a) and (b) we have plotted $ eB $-dependence of $ M $ at two different  values of  quark chemical potential with and without AMM of the quarks for $ T = 0 $ and $ 150  $ MeV. Since we have not used a sharp cutoff during numerical evaluation, an oscillatory behaviour of $M$ is observed.  These oscillations are related to the well known de Haas-van Alphen (dHvA) effect~\cite{Landau:1980mil} in the weak magnetic field regime and have also been observed  in Refs.~\cite{Fayazbakhsh1,Fayazbakhsh2,Sadooghi,Sadooghi1,Ebert:1998gx,Inagaki:2004ih,Noronha:2007wg,Fukushima:2007fc,Orlovsky:2014kva,Chaudhuri}. It occurs whenever the Landau levels pass the quark Fermi surface.  From Fig.~\ref{M_vs_eB} it is evident that, the dHvA oscillations get smeared out with the increase of the background magnetic field (as LLL dominates) in agreement with Ref.~\cite{Sadooghi,Chaudhuri}.	As expected from Figs.~\ref{MF0eBk_vs_T} (a) and (b), for a particular temperature there is an overall 	increase of $ M $ with $ eB $ when AMM of the quarks are not taken into consideration. On the other hand, inclusion of AMM 	leads to a reduction in $M$ with increasing $ eB $. These two phenomena indicates the occurrence of MC or IMC during the transition from broken to symmetry restored phase, as discussed earlier. 

\begin{figure}[h]
	\begin{center}
		\includegraphics[angle=-90, scale=0.4]{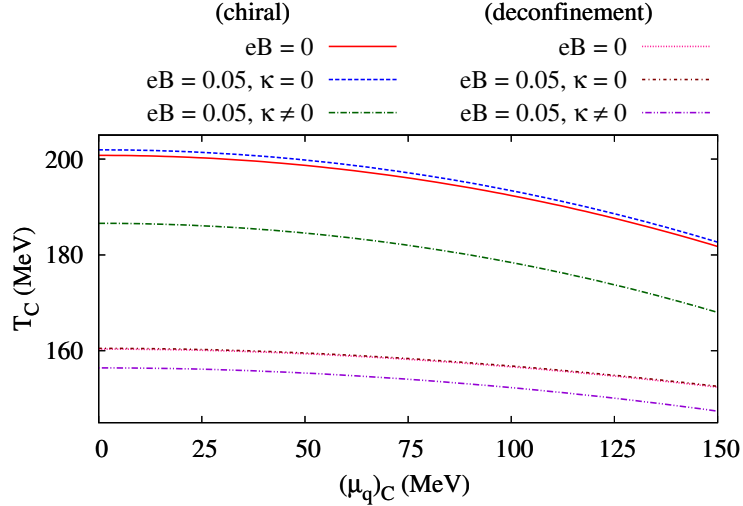}					
	\end{center}
	\caption{  $ T_C$-$ (\mu_q)_C $  phase diagram for chiral and deconfinement transition at different values of $ eB$ and $ \kappa  $. }
	\label{T_vs_muq_pd}
\end{figure}

We have used the peak positions of the 	 $ \chi_{MM} $ and 	$ \chi_{\F\Fb} $ susceptibilities to determine the phase boundaries in the $ T $-$ \mu_q $ plane following~\cite{SasakipNJL} and thus a direct correspondence between Figs.~\ref{chiMM_vs_T_eBk} (a) and (b) with the phase diagram of PNJL model, shown in Fig.~\ref{T_vs_muq_pd}, is evident. Notice that, with these parameters the boundary lines of  chiral symmetry restoration and deconfinement  transitions do not coincide~\cite{SasakipNJL,Costa}.  When we include only the  background magnetic field there is a slight increase in the  chiral symmetry restoration temperature for all values of quark chemical potential. On the contrary, consideration of non-zero AMM of the quarks decreases the chiral transition temperature throughout the whole range of $ (\mu_q)_C $ in the phase diagram. For deconfinement transition we observe that, magnetic field alone does not affect the deconfinement transition significantly. However, incorporation of finite AMM of the quarks results in a substantial decrease in the deconfinement transition temperature at each value of $ (\mu_q)_C $. 
\begin{figure}[h]
	\begin{center}
		\includegraphics[angle=-90, scale=0.6]{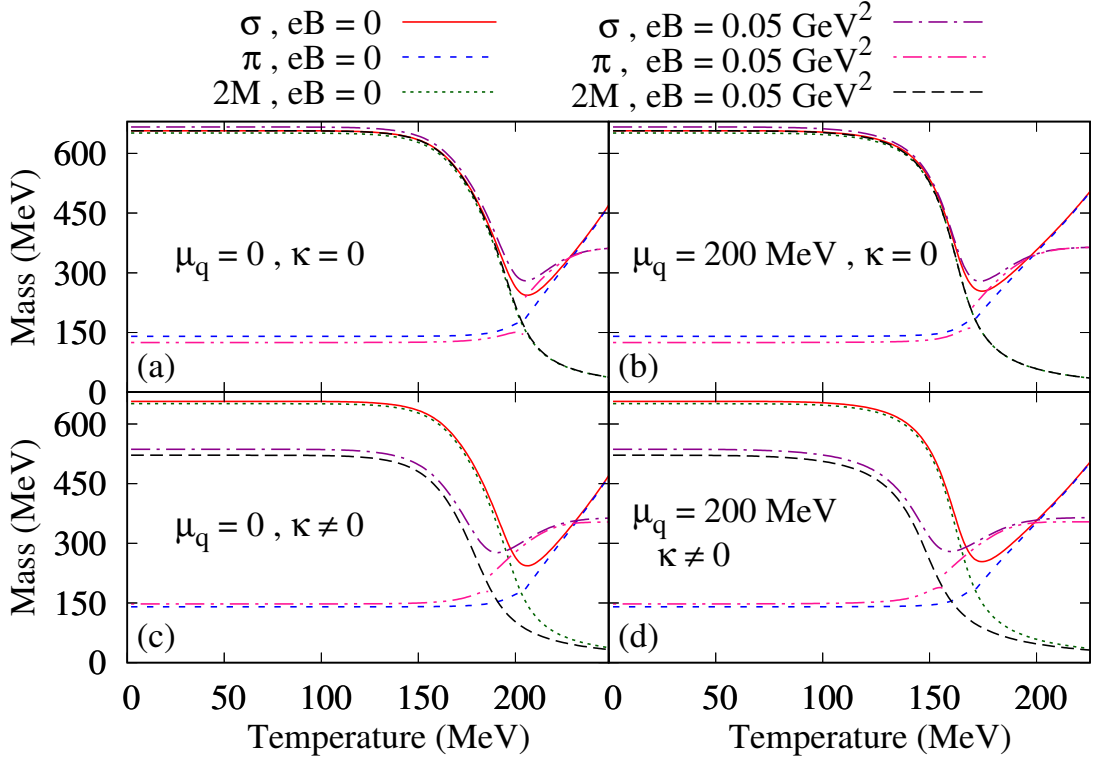}		
	\end{center}
	\caption{ Variation of scalar ($\sigma$) and neutral pseudo-scalar ($\pi^0$) meson masses as a function of temperature for different values of $\mu_q$ and $eB$ with and without considering the AMM of the quarks. The variation of twice the constituent quark mass has also been shown for comparison.} 
	\label{Mesons_vs_T}
\end{figure}
Now we turn our attention to the mesonic properties in the PNJL model under external magnetic field. In Fig.~\ref{Mesons_vs_T}, 
$m_\sigma$, $m_{\pi^0}$ and $2M$ have been plotted as a function of temperature. Figs.~\ref{Mesons_vs_T}(a) and (b) 
depict the variation of these quantities at $eB=0$ and $0.05$ GeV$^2$ without considering the AMM of the quarks 
at $\mu_q=0$ and $\mu_q=200$ MeV respectively, whereas Figs.~\ref{Mesons_vs_T}(c) and (d) depict the same for non-zero AMM of the quarks. 
It can be noticed that, at $T=0$ and $B=0$, all the mass graphs starts from the corresponding vacuum values. It can be seen that, $ m_\sigma $ remain almost unchanged up to $ T\backsimeq 100 $ MeV in all the cases, then decreases with the increase in temperature up to $ T_C $, attains a local minima around the transition temperature ($T\simeq T_c$) and then increases with the increase in $T$ at higher temperatures($T>T_C$).
On the contrary, $m_{\pi^0}$, 
being the mass of Goldstone boson associated with the chiral symmetry breaking, remains almost constant with the variation of 
temperature at the lower temperature ranges ($T<T_C$) in all the cases. Above the transition temperature ($T>T_C$), $m_{\pi^0}$ increases 
monotonically with the increase in $T$ and finally merges with  $m_\sigma$ as a consequence of the partial restoration of the 
chiral symmetry. It can also be observed that, $m_\sigma$ remains always greater that $2M$ in all the cases implying that $\sigma$ is always a resonant excitation whereas the value of $m_{\pi^0}$ is less than that of $2M$ at lower temperature range ($T\lesssim T_C$) indicating 
that $\pi^0$ is bound state at lower temperature. At higher temperatures ($T\gtrsim T_C$), $m_{\pi^0}>2M$ making $\pi^0$ 
a resonant excitation. The effect of increase of $\mu_q$ is seen to decrease the transition temperature for the chiral symmetry restoration 
and thus an overall shift of the mass graphs (keeping the qualitative nature same) towards the lower temperatures as can be noticed as one goes 
from Figs.~\ref{Mesons_vs_T}(a) and (c) to (b) and (d) respectively. When the AMM of the quarks is switched off, the change in the mass 
graphs with the increase in the external magnetic field is small as compared to the non-zero AMM case. At $\kappa=0$, $m_\sigma$ 
increases whereas the $m_{\pi^0}$ decreases with the increase in $eB$ in the lower temperature range. The scenario is completely reversed 
when the AMM of the quarks are switched on.  In this case, $m_\sigma$ decreases whereas $m_{\pi^0}$ increases with the increase in external 
magnetic field at low temperature. Similar results for scalar and pseudoscalar mass in presence of a background magnetic field without considering the finite values of AMM of the quarks have also been found in Ref.~\cite{Dumm:2020muy} using the non-local PNJL model. 
Moreover $m_{\pi^0}$ suffers a sudden jump~\cite{Avancini,MaoWang,Chaudhuri} 
at some particular temperature (for both  cases) which is a consequence of the dimensional 
reduction to (1+1)D due to external magnetic field.


%
\begin{figure}[h]
	\begin{center}
		\includegraphics[angle=-90, scale=0.35]{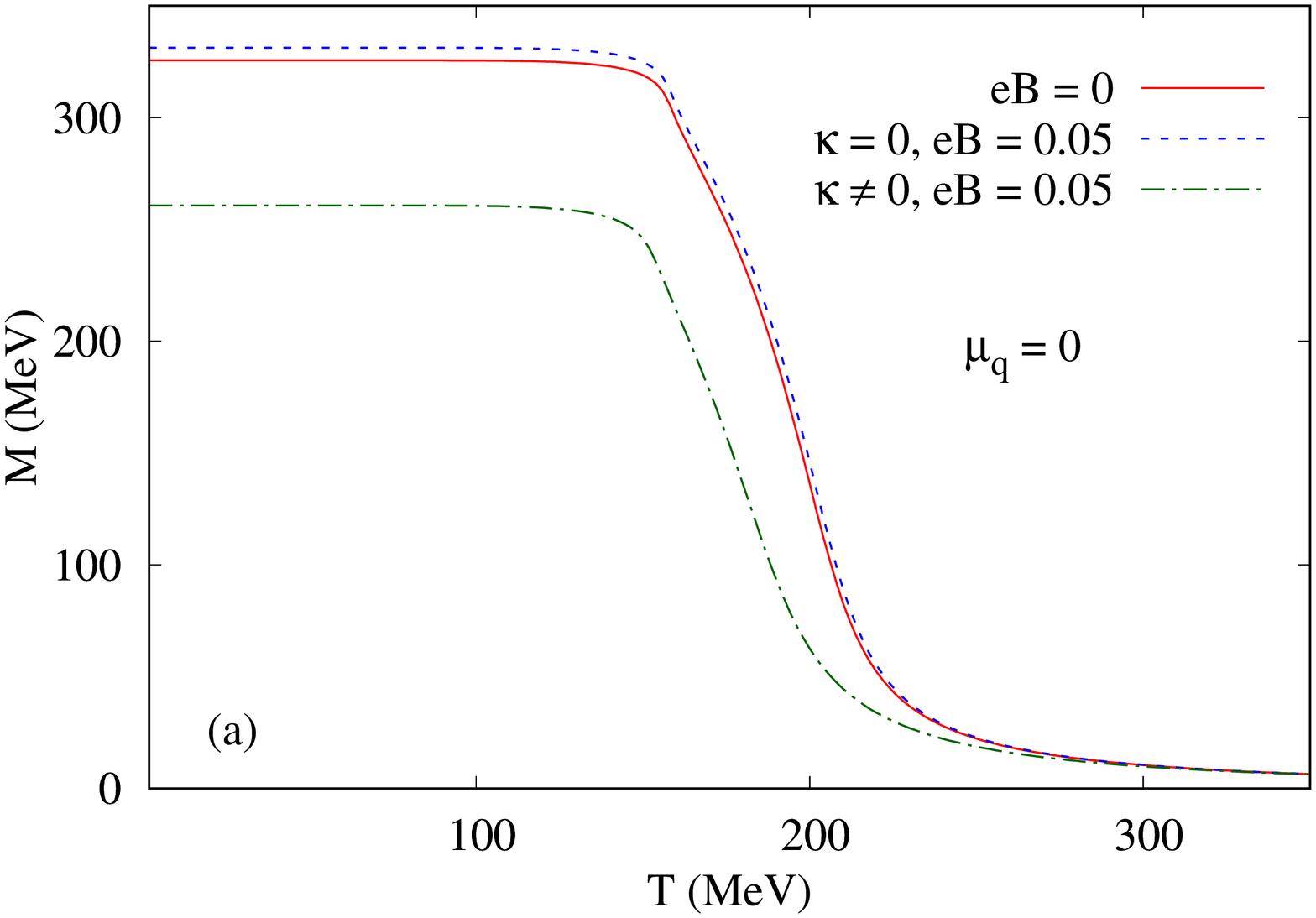}				\includegraphics[angle=-90, scale=0.35]{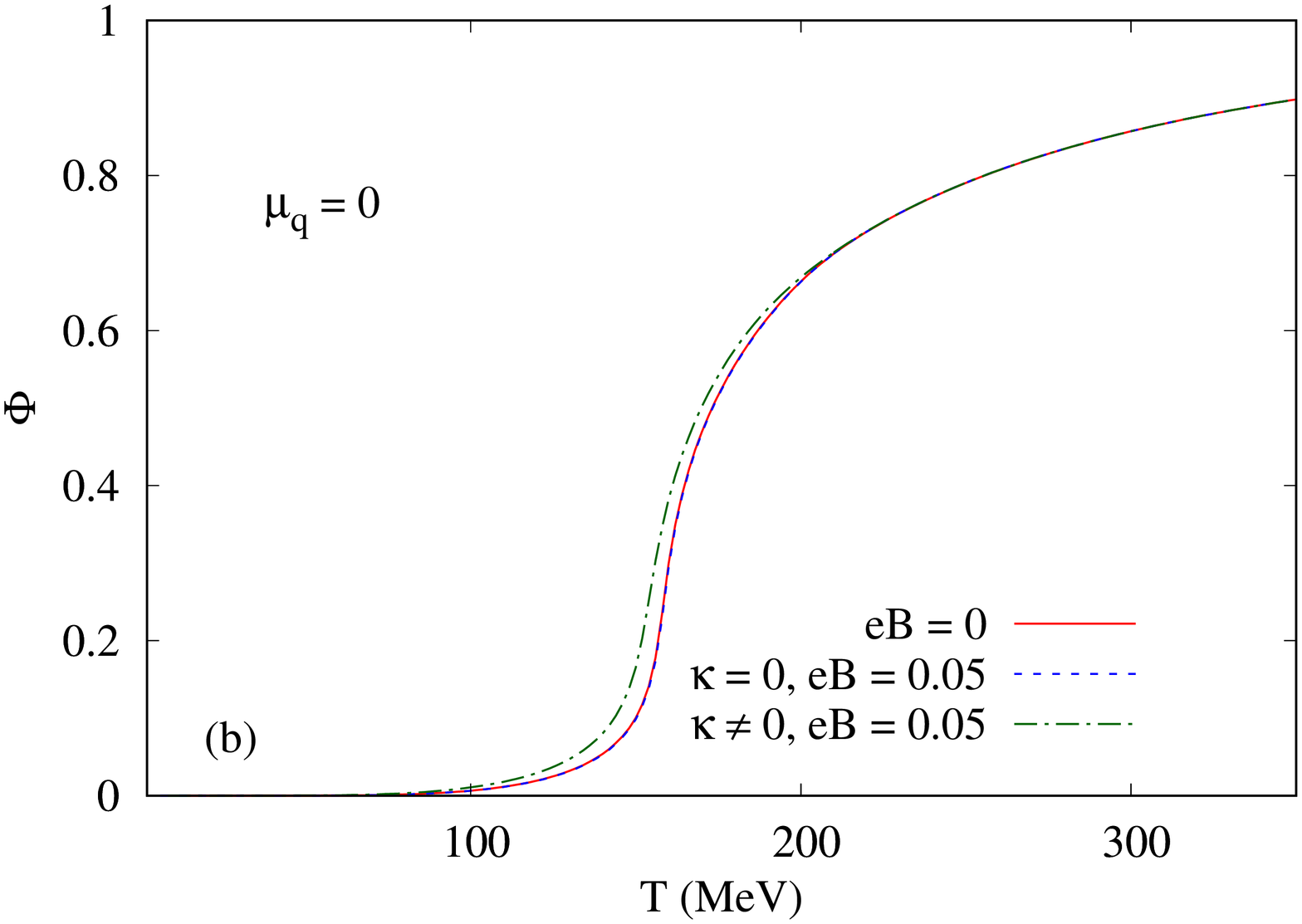}	\\
		\includegraphics[angle=-90, scale=0.35]{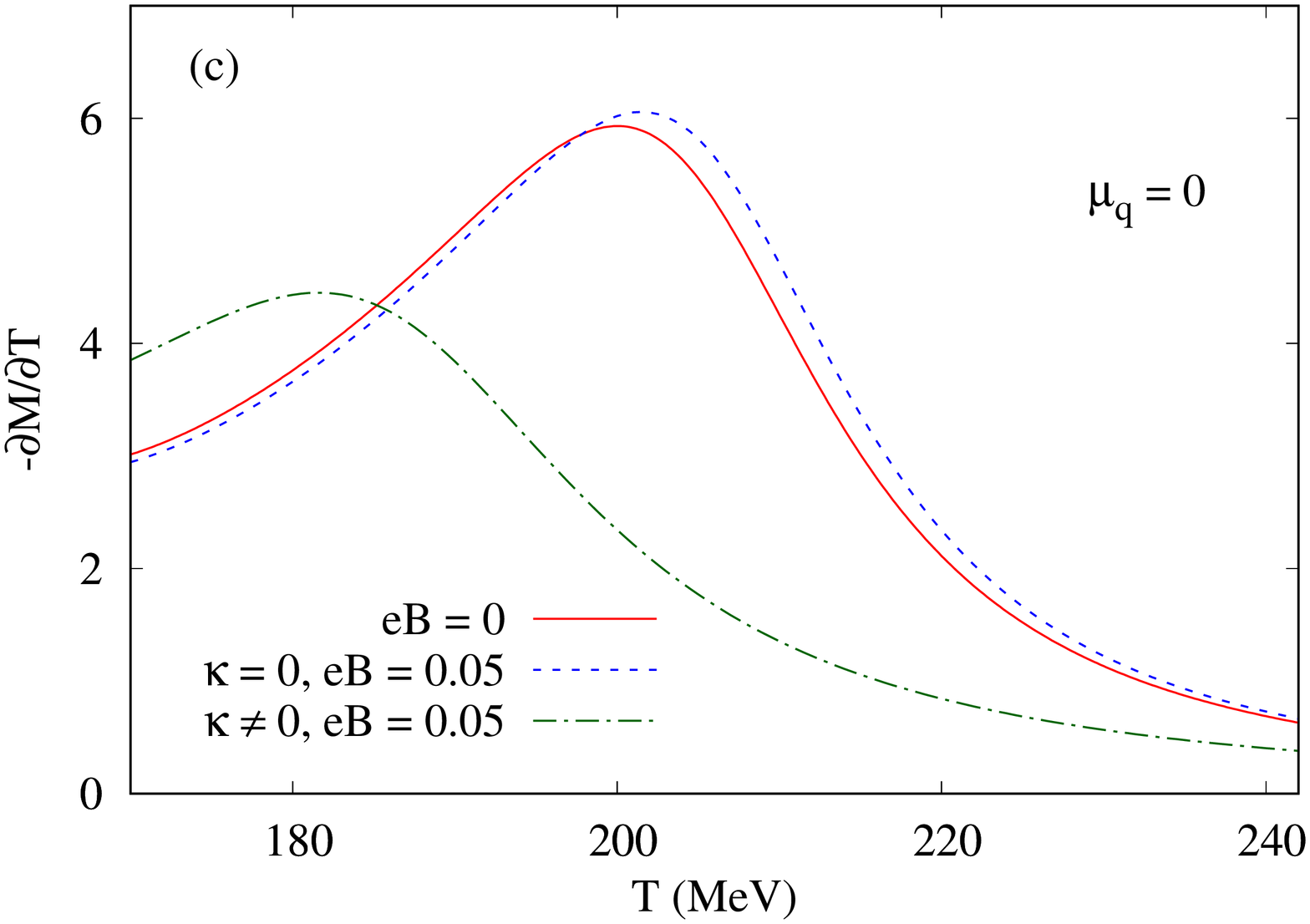}		
		\includegraphics[angle=-90, scale=0.35]{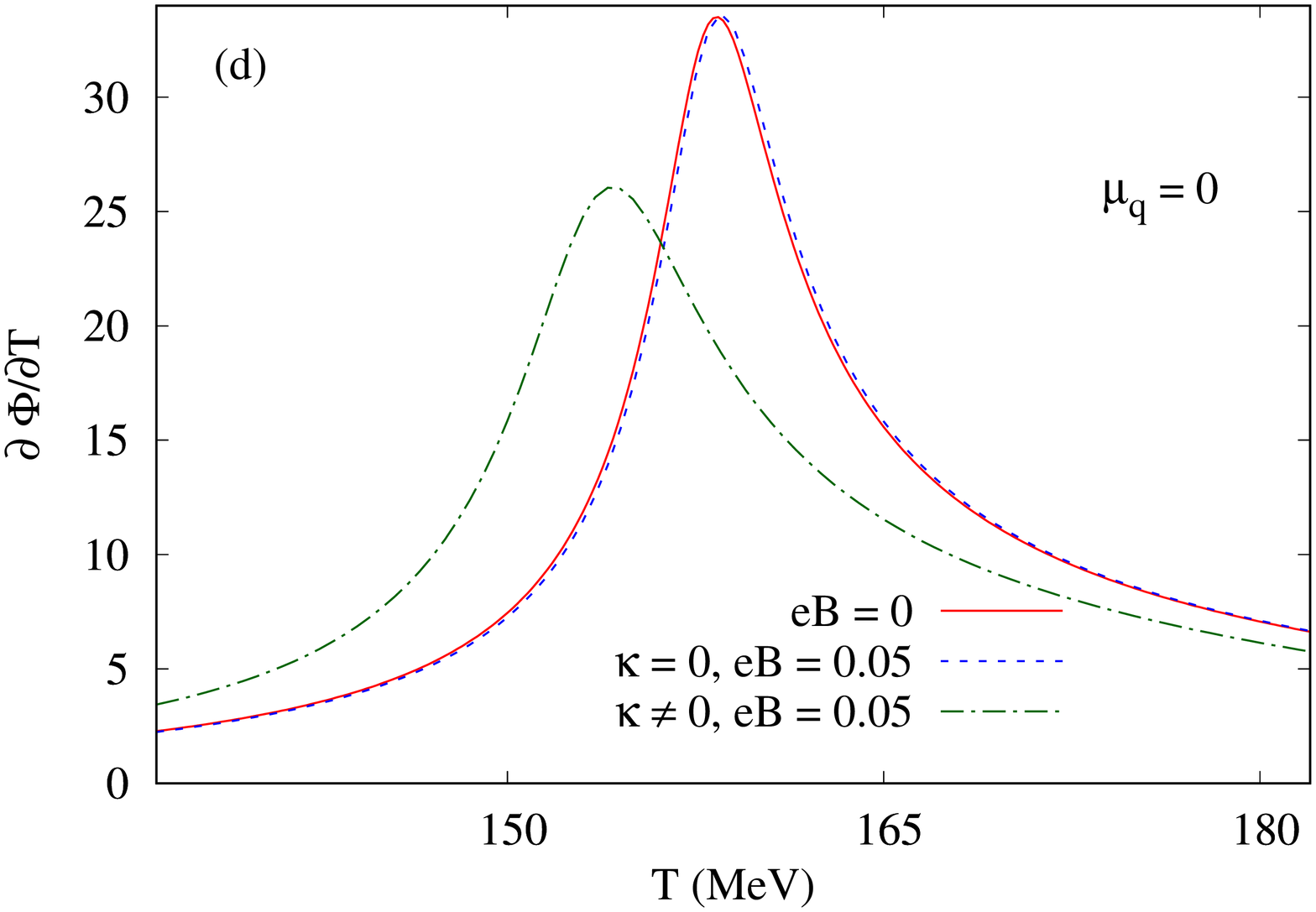}			
	\end{center}
	\caption{ Variation of $ M $, $ -\FB{\paroneder{M}{T}} $, $ \F $ and $ \FB{\paroneder{\F}{T}} $ as a function of $ T $ at $ \mu_q = 0 $ and different values of $ eB $  considering zero and non-zero values of the AMM of the quarks using the Polyakov loop potential defined in Eq.~\eqref{polyakov_potential_2nd}.}
	\label{Comp_pot}
\end{figure}

It may be noted that, the IMC in chiral and deconfinement transitions due to inclusion of the AMM of the quarks, as we have seen while discussing Fig.~\ref{MF0eBk_vs_T}, is not particular to the choice of Polyakov loop potential (Eq.~\eqref{polyakov_potential}). In the following we have considered another form of Polyakov  loop potential used frequently in the literature~\cite{Ratti3,Fukushima_CME_pNJL,Fukushima_pNJLPD,Gatto,Gatto2}:
 \begin{eqnarray}\label{polyakov_potential_2nd}
 \frac{\mathcal{U}\FB{ \Phi,\bar{\Phi} ;T}}{T^4} = -\frac{a(T)}{2}  \bar{\Phi} \Phi  + b(T) \ln \TB{1 - 6\bar{\Phi}\Phi  + 4 \FB{ \bar{\Phi}^3 -  \Phi^3 } - 3 \FB{\bar{\Phi}\Phi}^2 }
 \end{eqnarray} 
 where 
 \begin{eqnarray}
 a(T) = a_0 + a_1 \FB{\frac{T_0}{T}}+ a_2 \FB{\frac{T_0}{T}}^2 ;~~~~~~~~ b(T) =  b_3 \FB{\dfrac{T_0}{T}}^3.
 \end{eqnarray} 
 All the parameters are defined in~\cite{Ratti3}. Using this form of the potential in Figs.~\ref{Comp_pot}  (a)-(d) we have shown the variation of constituent quarks mass $ M $ and the expectation value of Polyakov loop $ \Phi $ and their derivatives as a function of temperature at $ \mu_q = 0 $. Comparing this with  Fig.~\ref{MF0eBk_vs_T}, one can see that the results are qualitatively the same. Both the chiral as well as deconfinement transitions show IMC when finite values of the AMM  of the quarks are taken into consideration. An opposite effect is observed when the AMM of the quarks are switched off, which can be identified as MC.

\section{Summary \& Conclusion}\label{sec:conclusion}
In the present work, we have studied the 2-flavor PNJL model at finite temperature and baryonic density in presence of arbitrary external magnetic field with the inclusion of AMM of the quarks. The variation of constituent quark mass ($ M $) and the traced Polyakov loop ($ \F $) as a function of $ T $ and $ \mu_q $ is obtained by solving the coupled gap equations. Examining $ M $ as a function $ T $ for a given value of $ \mu_q $,  the transition temperature from chiral symmetry broken to restored phase is observed to increase with the increase in external magnetic field owing to the enhancement of quark anti-quark condensate. This observation is further confirmed by studying the $ T $-dependence of  the quantity $ -\FB{\paroneder{M}{T}} $ where the peak of the curve, which can be identified as chiral transition temperature ($ T_C^\chi$) is found to move towards higher values of temperature as the magnetic field is increased. This phenomena can be classified as MC. On the contrary, when we include finite values of the AMM of the quarks in presence of external magnetic field of same strength, $ M $ is found to be smaller as compared to the zero AMM case (for all values of $ T $). $ T_C^\chi $ for non-zero AMM case decreases with the increase in $eB$ which is also evident if one considers the plot of $ -\FB{\paroneder{M}{T}} $ as a function of $ T $ whose peak is found to shift towards lower temperature values indicating IMC. But, when AMM of the quarks are switched off, the external magnetic field is found to affect the $ T $-dependence of $ \F $ marginally.  However, when the AMM of the quarks are considered, the temperature for transition from confined to deconfined phase ($ T_C^d $) is observed to decrease with the increase in external magnetic field. A similar conclusion about the effect of inclusion and exclusion of  AMM in presence of background magnetic field in the behaviour of  $  T_C^\chi$  and $ T_C^d $ is evident from studying $\mu_q$-dependence of $ M $ and $ \F $. Evidence for the occurrence of a possible quarkyonic phase, i.e., the phase in which	quarks remain confined ($ \Phi \lesssim 0.4$) even though chiral symmetry has been restored, is found at low $ T $ and high $ \mu_q $. Searching for this phase is one of the important goals of NICA~\cite{nica}. Interestingly when we consider finite value of AMM of the quarks the presence of the quarkyonic phase may be possible even at higher values of $ T $. 

Several thermodynamic quantities such as scaled pressure, entropy and energy density are calculated at zero quark chemical potential and it is observed that they behave similarly as all the three curves increase sharply in the vicinity of the phase transition owing to the liberation of degrees of freedom and eventually saturate (approaching the corresponding Stefan-Boltzmann limits). For all values of temperature as well as finite values of background magnetic field with or without including the AMM of the quarks, all the thermodynamic variables previously mentioned are observed to vary smoothly with temperature indicating the fact that the associated phase transition is a crossover. Although for finite values of AMM of the quarks, we find that the transition occurs at lower $ T $ values. Reduced quark number density is studied at different values of $ \mu_q $ and it is observed that with increasing temperature it increases monotonically, attains a local maxima around the transition temperature and finally decreases slowly with increasing temperature. Inclusion of AMM of the quarks causes two noticeable  differences.  Firstly, when  AMM of the quarks is turned on, because of the finite values of $ \F $ and lower values of $ M $ compared to the zero AMM case,  the dominance of three-quark states survives for lower range of $ T  $ values. This results in a sharp increase in $ n_q/T^3 $ at lower values of $ T $.  Secondly,  even at high values of $ T $, $ M $ has sufficiently low magnitude in case of finite AMM of the quarks which leads to larger suppression in $ n_q $ compared to the cases when AMM of the quarks are ignored. Similar features are also reflected in other thermodynamic quantities such as the specific heat ($ C_V $), velocity of sound squared ($ c_s^2 $) and quark number susceptibility ($ \chi_q $).

Next using $ \chi_{MM} $ and $ \chi_{\F\Fb} $, which are the susceptibilities related to $ M $ and $ \F $ respectively, we evaluate the chiral ($ T_C^\chi $) and deconfinement ($ T_C^d $) transition temperatures. With our choice of parameters, $ T_c^\chi $ and $ T_C^d $ do not coincide at vanishing quark chemical potential. The peaks of $ \chi_{MM} $ and $ \chi_{\F\Fb} $ are then used to draw the $ T_C $-$ (\mu_q)_C$ phase diagram for both chiral and deconfinement phase  transitions for the three cases previously mentioned. We find that switching on the background magnetic field results in a slight increase in $ T_C^\chi$ for the whole range of $ (\mu_q)_C$ values, however, $ T_C^d $ remains nearly unaltered. On the other hand, while considering finite AMM of the quarks in presence of background magnetic field we observe that both the chiral and deconfinement transitions occur at lower values of temperature throughout the whole range of $ (\mu_q)_C $ in the phase diagram.

The masses of the scalar ($ \sigma $) and  neutral pseudoscalar ($ \pi^0 $) mesons have been evaluated considering a hot and dense magnetized medium using the RPA in the PNJL model. For this, both the AMM of the quarks as well as infinite number of quark Landau levels are taken into consideration in the analytical and numerical  calculations so that the results are valid for an arbitrary strength of the external magnetic field. It is observed that, $m_\sigma$ at finite values of external magnetic field noticeably decreases while considering the AMM of the quarks as compared to the zero AMM case. On the contrary, the $m_{\pi^0}$ remains almost constant (close to the vacuum value $\simeq 140$ MeV) at the lower temperature range irrespective of the consideration of the AMM thus maintaining the signature of the Nambu-Goldstone boson. 

We end by noting that in a theory with massless charged fermions it is not possible to find an anomalous magnetic moment using Schwinger’s perturbative approach~\cite{Ferrer:2009nq} so that the linear-$ B $ ansatz~\cite{Schwinger:1948iu} is not valid anymore. Presence of an AMM would break the chiral symmetry of the massless theory which is protected against any perturbatively generated breaking term. However, massless charged fermions in the presence of a 	magnetic field can acquire a dynamical magnetic moment~\cite{Mao:2018jdo,Ferrer:2009nq} which goes to zero in the chiral symmetry restored phase~\cite{Mao:2018jdo}. Since the chiral limit is achieved in this phase and $\kappa_f\to 0$, the gapless nature of the LLL is maintained. Since we have considered a constant value of AMM, this feature is absent here. 
A dynamic evaluation of AMM of the quarks incorporating the essential features will be presented elsewhere.


\section*{Acknowledgments}
The authors were funded by the Department of Atomic Energy (DAE), Government of India.


\appendix
\section{ DOUBLE DERIVATIVES OF $ \Omega $  WITH RESPECT TO $ M, \F $ AND $ \overline{\Phi} $  }\label{double_der}
From Eq.~\eqref{Omega_PNJL} we get 
\begin{eqnarray}
\paroneder{\Omega}{M} &=& \frac{M- m}{G} - 3 \sum_{n,f,s} \frac{\MB{q_f B}}{2\pi^2 } \intzinf dp_z \frac{M}{\omega_{nfs} } \FB{ 1 - \frac{ s\kappa_f q_f B}{M_{nfs}}  } \TB{ \frac{}{}1-  f^+ \fnppbT - f^- \fnppbT}.\label{delOmegadelM}
\end{eqnarray}
Following relations can be used to arrive at the above result
\begin{eqnarray}
\paroneder{\omega_{nfs}}{M} &=& \frac{M}{\omega_{nfs} } \FB{1 - \frac{s\kappa_f q_f B}{M_{nfs}} }, \\
\paroneder{\ExpFac{n}{\mp}}{M} &=& -\frac{n\beta M}{\omega_{nfs}} \AMMbracket \ExpFac{n}{\mp }, \\
\paroneder{\ln g^{(+)}}{M} &=& -\frac{3 \beta M}{\omega_{nfs}} \AMMbracket f^+ \fnppbT,\\
\paroneder{\ln g^{(-)}}{M} &=& -\frac{3 \beta M}{\omega_{nfs}} \AMMbracket f^- \fnppbT.
\end{eqnarray}
Note that in Eq.~\eqref{delOmegadelM} the medium independent term has to be regularized by introducing a field dependent cutoff (see~\cite{Chaudhuri} for details):
\begin{equation}
\Lambda_z = \sqrt{\Lambda^2 - (2n + 1-s ) \MB{q_f B} + 2M_{nfs} s\kappa_f q_f B - (\kappa_f q_f B)^2   }. 
\end{equation}
So the regularized version of Eq.~\eqref{delOmegadelM} is 
\begin{eqnarray}
\paroneder{\Omega}{M} &=& \frac{M- m}{G}-  3 \sum_{n,f,s} \frac{\MB{q_f B}}{2\pi^2 } \int_0^{\Lambda_z} dp_z \frac{M}{\omega_{nfs} } \AMMbracket + 3 \sum_{n,f,s} \frac{\MB{q_f B}}{2\pi^2 }   \intzinf dp_z \frac{M}{\omega_{nfs} } \FB{ 1 - \frac{ s\kappa_f q_f B}{M_{nfs}}  }\nn \\ && ~~~~~~~~~~~~~~~~~~~~~~~~~~~\times  \TB{ \frac{}{} f^+ \fnppbT + f^- \fnppbT}\label{regdOdM} .
\end{eqnarray}
Now to evaluate the second derivative with respect to $ M $, the following relations will be useful:
\begin{eqnarray}
\frac{M}{\omega_{nfs} }\AMMbracket \bigg|_{p_z = \Lambda_z} &=& \frac{M^ 2}{\Lambda_z\sqrt{\Lambda^2 + M^2 }} \frac{s\kappa_f q_f B}{M_{nfs}} \AMMbracket \\
\paroneder{\Lambda_z }{M}   &=& \frac{s\kappa_f q_f B}{\Lambda_z} \frac{M}{M_{nfs}}  , 
\end{eqnarray}
\begin{eqnarray}
\frac{\partial}{\partial M} {\AMMbracket} &=& \frac{1}{\omega_{nfs}}\AMMbracket - \frac{M^2}{\omega_{nfs}^3} \AMMbracket^2 + \frac{M^2 s\kappa_f q_f B}{\omega_{nfs} M_{nfs}^3}.
\end{eqnarray}
Thus we can finally write
\begin{eqnarray}
\partwoder{\Omega}{M } &=& \frac{1}{G} - 3 \sum_{n,f,s} \frac{\MB{q_f B}}{2\pi^2 } \int_{0}^{\Lambda_z }  dp_z \TB{\frac{1}{\omega_{nfs}}\AMMbracket - \frac{M^2}{\omega_{nfs}^3} \AMMbracket^2 + \frac{M^2 s\kappa_f q_f B}{\omega_{nfs} M_{nfs}^3}}\TB{\frac{ }{}  1 - f^+ \fnppbT- f^- \fnppbT } \nn \\ && - 3 \sum_{n,f,s} \frac{\MB{q_f B}}{2\pi^2 } \frac{M^ 2}{\Lambda_z\sqrt{\Lambda^2 + M^2 }} \frac{s\kappa_f q_f B}{M_{nfs}} \AMMbracket - \frac{3}{T}\eBprefac\frac{M^2}{\omega_{nfs}^2}\AMMbracket^2\TB{ \frac{\ExpFac{ }{- }}{g^{(+)}} \nn \right. \\&& \left.\times  \SB{ \FB{\frac{}{ }\F + 4 \Fb \ExpFac{ }{- }+ 3 \ExpFac{2}{-}  }-3f^+\fnppbT \FB{\F + 2\Fb \ExpFac{ }{-}  + \ExpFac{ 2}{-}  }} \nn \right. \\ && \left.  + \SB{ \frac{}{ }\F \leftrightarrow \Fb; \mu_q \rightarrow -\mu_q } },
 \end{eqnarray}
\begin{eqnarray}
\parmultwoder{\Omega}{\Phi}{M} &=& 3 \eBprefac \frac{M}{\omega_{nfs} } \AMMbracket \TB{\frac{\ExpFac{ }{-}}{g^{(+)}} -\frac{3f^+\fnppbT}{g^{(+)}} \ExpFac{ }{-}  \nn  \right. \\ && \left. + \frac{2\ExpFac{ 2}{+}}{g^{(-)}} -\frac{3f^-\fnppbT}{g^{(-)}} \ExpFac{2 }{+}  } ,\\
\parmultwoder{\Omega}{\Fb }{M} &=& \parmultwoder{\Omega}{\Phi}{M} \SB{\frac{}{} \F \leftrightarrow \Fb ; \mu_q \rightarrow -\mu_q  },
\end{eqnarray}
\begin{eqnarray}
\partwoder{\Omega}{\F} &=& \FB{-b_3 \F + \frac{b_4}{2} \Fb^2}T^4 + 9 T \eBprefac \TB{\frac{\ExpFac{2}{-}}{{g^{(+)}}^2  }  + \frac{\ExpFac{4}{+}}{{g^{(-)}}^2  }  },\\
\partwoder{\Omega}{\Fb} &=&\partwoder{\Omega}{\F}  \SB{\frac{}{} \F \leftrightarrow \Fb ; \mu_q \rightarrow -\mu_q  },
\end{eqnarray}
\begin{eqnarray}
	\parmultwoder{\Omega}{\F}{\Fb} &=& \FB{\frac{- b_2(T)}{2} + b_4 \Fb \F   }T^4 +9 T \eBprefac \TB{\frac{\ExpFac{3}{-}}{{g^{(+)}}^2  }  + \frac{\ExpFac{3}{+}}{{g^{(-)}}^2  }  }.
\end{eqnarray}

\section{$ T $-DERIVATIVES OF $ M, \F, \overline{\Phi} $} \label{T_der} 
We have
\begin{eqnarray}
\dfrac{\partial }{\partial T} \TB{\ExpFac{n}{\mp }} &=& n \ExpFac{n}{\mp } \TB{ \frac{\omega_{nfs} \mp \mu_q }{T^2}  - \frac{M}{\omega_{nfs} } \AMMbracket \FB{\paroneder{M}{T}} } ,
\end{eqnarray}
\begin{eqnarray}
\paroneder{f^+\fnppbT}{T}&=&  \FB{\paroneder{f^+}{M}}\FB{\paroneder{M}{T}} +  \FB{\paroneder{f^+}{\F}}\FB{\paroneder{\F}{T}} +\FB{  \paroneder{f^+}{\Fb}}\FB{\paroneder{\Fb}{T}} + A^+_{M,T} 	
\end{eqnarray}
where
\begin{eqnarray}
\paroneder{f^+}{M } &=& - \frac{M}{\omega_{nfs}} \AMMbracket \frac{\ExpFac{ }{-}}{g^{(+)}} \TB{ \FB{\frac{}{ }\F + 4 \Fb \ExpFac{ }{- }+ 3 \ExpFac{2}{-}  } \nn \right. \\&&\left. -3f^+\fnppbT \FB{\F + 2\Fb \ExpFac{ }{-}  + \ExpFac{ 2}{-}  } } ,\\
\paroneder{f^+}{\F } &=& \frac{\ExpFac{ }{-}}{g^{(+)}} -\frac{3f^+\fnppbT}{g^{(+)}} \ExpFac{ }{-}, \\
\paroneder{f^+}{\Fb } &=& \frac{2\ExpFac{ 2}{-}}{g^{(+)}} -\frac{3f^+\fnppbT}{g^{(+)}} \ExpFac{2 }{-},
\end{eqnarray}

\begin{eqnarray}
A^+_{M,T} &=& \frac{\ExpFac{}{-} }{{g^{(+)}}^2} \frac{\FB{\omega_{nfs} -\mu_q }}{T^2} \SB{  \F  + 4\Fb \ExpFac{ }{-} + 3\FB{ 1 + \Fb \F  }\ExpFac{2}{-}  \nn \right. \\ && \left.+4\F \ExpFac{3}{-}  + \Fb \ExpFac{4}{-}  } ,
\end{eqnarray}
\begin{eqnarray}
\paroneder{f^-\fnppbT}{T}&=&\paroneder{f^+\fnppbT}{T} \SB{ \F \leftrightarrow \Fb;\frac{}{} \mu_q \rightarrow -\mu_q } .
\end{eqnarray}
Now using the above relations and results given in Appendix~\ref{double_der}, $ T $-derivatives of the gap equations of $ M, \F $ and $ \Fb $ can be calculated starting from Eqs.~\eqref{Gap_M},~\eqref{Gap_p} and \eqref{Gap_pb}. The expression can be written in a matrix form in the following way:
\begingroup
\renewcommand*{\arraystretch}{1.6}
\begin{eqnarray}
\begin{bmatrix}
C_{MM}   & C_{M\F} & C_{M\Fb}   \\ C_{\F M}   & C_{\F \F } & C_{\F\Fb}   \\ C_{\Fb M}   & C_{\Fb \F } & C_{\Fb \Fb}  
\end{bmatrix} 
\begin{bmatrix}
\frac{1}{\Lambda} \paroneder{M}{T}  \\  \paroneder{\F}{T} \\  \paroneder{\Fb}{T} 
\end{bmatrix} = \begin{bmatrix}
\frac{T}{\Lambda^2} A_{M,T} \\ \frac{T^2}{\Lambda^3} A_{\F,T} \\ \frac{T^2}{\Lambda^3} A_{\Fb,T} 
\end{bmatrix} 
\end{eqnarray} 
\endgroup
 where
 \begin{eqnarray}
 A_{M,T} &=&-\frac{3}{T^4} \eBprefac \frac{M}{\omega_{nfs} }\AMMbracket \TB{  \frac{\ExpFac{}{-} }{{g^{(+)}}^2} \FB{\omega_{nfs} -\mu_q }\SB{  \F  + 4\Fb \ExpFac{ }{-} + 3\FB{ 1 + \Fb \F  }\ExpFac{2}{-}  \nn \right. \right.\\ && \left. \left.+4\F \ExpFac{3}{-}  + \Fb \ExpFac{4}{-}  }  + \replacement },
 \end{eqnarray}
\begin{eqnarray}
A_{\F,T} &=&-	\frac{T}{2}\paroneder{b_2(T)}{T} \Fb - \frac{9}{T^3}\eBprefac \FB{ \frac{\ExpFac{}{-}}{g^{(+)}} + \frac{\ExpFac{2}{+}}{g^{(-)}}  }\nn \\ && + \frac{3}{T^4} \eBprefac \TB{ \frac{\ExpFac{}{-} }{{g^{(+)}}^2}  \FB{\omega_{nfs} -\mu_q }  \SB{  1 - 3 \Fb \ExpFac{2}{-} - 2 \ExpFac{3}{-}   } \nn \right. \\ && \left.  + \frac{\ExpFac{2}{+} }{{g^{(-)}}^2} \FB{\omega_{nfs} +\mu_q } \SB{  2+ 3 \Fb \ExpFac{}{+} -  \ExpFac{3}{+}   }   }\\
{\rm and }\nn \\
A_{\Fb,T} &=&A_{\F,T}\replacement.
\end{eqnarray}
During this calculation we have put a combination of $ T $ and $ \Lambda $ with several quantities to make sure we get matrix with dimensionless co-efficients as introduced in Sec.~\ref{TD_PNJL}.

\section{ $\mu_q $-DERIVATIVES OF $ M,\F ,\overline{\Phi} $  }\label{mu_der}

Similar matrix form can also be written for  $ \mu_q $-derivatives of the gap equations as shown below
%
%
\begingroup
\renewcommand*{\arraystretch}{1.6}
\begin{eqnarray}
\begin{bmatrix}
C_{MM}   & C_{M\F} & C_{M\Fb}  \\ C_{\F M}   & C_{\F \F } & C_{\F\Fb}    \\ C_{\Fb M}   & C_{\Fb \F } & C_{\Fb \Fb}  
\end{bmatrix} 
\begin{bmatrix}
\frac{1}{\Lambda} \paroneder{M}{\mu_q}\\   \paroneder{\F}{\mu_q} \\   \paroneder{\Fb}{\mu_q} 
\end{bmatrix} = \begin{bmatrix}
\frac{T}{\Lambda^2} A_{M,\mu_q} \\  \frac{T^2}{\Lambda^3} A_{\F,\mu_q} \\  \frac{T^2}{\Lambda^3} A_{\Fb,\mu_q} 
\end{bmatrix} 
\end{eqnarray} 
\endgroup
where
\begin{eqnarray}
A_{M,\mu_q} &=&-\frac{3}{T^3} \eBprefac \frac{M}{\omega_{nfs} }\AMMbracket \TB{  \frac{\ExpFac{}{-} }{{g^{(+)}}^2} \SB{  \F  + 4\Fb \ExpFac{ }{-} + 3\FB{ 1 + \Fb \F  }\ExpFac{2}{-}  \nn \right. \right.\\ && \left. \left.+4\F \ExpFac{3}{-}  + \Fb \ExpFac{4}{-}  }  - \replacement },
\end{eqnarray}
\begin{eqnarray}
A_{\F,\mu_q} &=& \frac{3}{T^3} \eBprefac \TB{ \frac{\ExpFac{}{-} }{{g^{(+)}}^2}\SB{  1 - 3 \Fb \ExpFac{2}{-} - 2 \ExpFac{3}{-}   } \nn \right. \\&&\left. + \frac{\ExpFac{2}{+} }{{g^{(-)}}^2}\SB{  2+ 3 \Fb \ExpFac{}{+} -  \ExpFac{3}{+}   }   },\\
A_{\Fb,\mu_q} &=&A_{\F,\mu_q }\replacement.
\end{eqnarray}

\bibliographystyle{apsrev4-1}
\bibliography{Nilanjan}

\end{document}